\documentclass{article}

 \usepackage[main, final]{neurips_2026}

\usepackage{graphicx} 
\usepackage{amsmath, amssymb}
\usepackage{geometry}
\usepackage{hyperref}
\usepackage{booktabs}
\usepackage{eso-pic} 
\usepackage{xcolor}
\usepackage[table]{xcolor}
\renewcommand{\arraystretch}{1.}
\usepackage{amsthm}
\usepackage{physics}
\usepackage{subcaption}
\usepackage{multirow}
\usepackage{amsthm}
\usepackage{mathtools}
\usepackage{fontawesome5}
\usepackage{enumitem}
\setlist[itemize]{noitemsep, topsep=0pt}
\usepackage{algpseudocode}
\newtheorem{theorem}{Theorem}[section]

\newsavebox{\logo}
\savebox{\logo}{\includegraphics[width=1.3in]{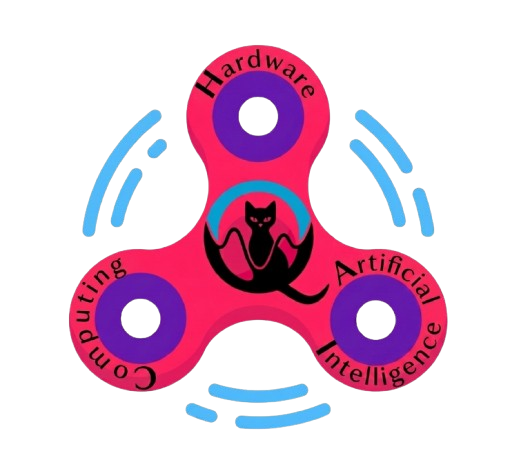}}%
\AddToShipoutPictureFG*{%
	\put(
	\LenToUnit{\dimexpr\paperwidth-1.6in},
	\LenToUnit{\dimexpr\paperheight-1.65in}  
	){\usebox\logo}%
}

\usepackage[ruled, linesnumbered]{algorithm2e}
\newcommand{\htwoo}{\ensuremath{\texttt{H}_2\texttt{O}}}

\newcommand{\behtwo}{\ensuremath{\texttt{BEH}_2}}

\SetKwProg{Proc}{Procedure}{}{end}
\SetKwFunction{FPush}{Push}
\SetKwFunction{FCompute}{ComputeEpisodePriorities}
\SetKwFunction{FSample}{Sample}
\SetKwFunction{FUpdate}{UpdatePriorities}
\SetKwFunction{FTrain}{TrainStep}
\SetKwFunction{FClip}{clip}

\newtheorem{proposition}{Proposition}
\newtheorem{assumption}{Assumption}        
\newtheorem{lemma}[theorem]{Lemma}

\title{Replay-buffer engineering for noise-aware quantum circuit optimization}

\vspace{-10pt}
\author{Akash Kundu$^{1, 2}$\thanks{Corresponding author.} \quad Sebastian Feld$^{1,2}$ \\
$^1$Delft University of Technology, Delft, The Netherlands\\
$^2$Quantum Computing Division, QuTech, The Netherlands \\
\texttt{\{A.kundu, S.Feld\}@tudelft.nl} \\\\
\faGithub\ \href{https://github.com/Aqasch/replay_buffer_eng_for_quantum_optimization}{Code on GitHub}
}

\begin{document}
\maketitle

\vspace{-12pt}
\begin{abstract}
Deep reinforcement learning for quantum circuit optimization faces three bottlenecks: replay buffers that overlook temporal difference (TD) target reliability, curriculum-based architecture search requiring a full quantum-classical evaluation after every edit, and the discard of noiseless trajectories when retraining under hardware noise. We address these limitations by treating replay as a central algorithmic lever. We introduce ReaPER+, an annealed replay rule that transitions from TD-error prioritization to reliability-aware sampling as value estimates mature. ReaPER+ achieves up to 4x higher sample efficiency than fixed PER, ReaPER, and uniform replay, while matching prior on-policy solution quality with up to 32x fewer interactions At 12 qubits, fixed ReaPER reaches the lowest energy error in the fewest steps, while PER and uniform replay find more compact circuits at higher error. On tasks scaling to 20 qubits, ReaPER+ retains its advantage, demonstrating that reliability-aware annealing extends beyond small-system benchmarks. LunarLander-v3 confirms that the ReaPER+ is domain-agnostic, it improves success rates by up to 26.8\% over PER and 21.8\% over fixed ReaPER, with a 3\% AUC gain over both. We further introduce OptCRLQAS, which amortizes quantum-classical evaluations across multiple architectural edits, reducing training wall-clock time by up to 67.5\% on 12-qubit without degrading solution quality. Finally, lightweight replay-buffer transfer warm-starts noisy optimization from noiseless trajectories, without weight transfer or $\epsilon$-greedy pretraining, reducing steps to chemical accuracy by 85-90\% and final energy error by up to 90\% relative to from-scratch learning. Transfer gains increase with system size. Together, these results establish experience storage, sampling, and transfer as decisive levers for sample efficient, noise-aware quantum circuit optimization.
\end{abstract}

\section{Introduction}
\vspace{-8pt}
Optimization is one of the most important application areas of quantum computing~\cite{abbas2024challenges,farhi2014quantum}, both as a target problem class and as an internal algorithmic task in the design and execution of quantum algorithms. On current quantum hardware, quantum computation is realized through sequences of elementary logic gates arranged into quantum circuits~\cite{devoret2013superconducting,croot2025enabling,kjaergaard2020superconducting}, and their depth, gate count, and hardware compatibility of these circuits directly determines what can be achieved in practice~\cite{kandala2017hardware,kim2023evidence}.

This makes circuit optimization a central challenge across both near-term and fault-tolerant regimes. In the near term~\cite{preskill2018quantum,bharti2022noisy}, limited coherence, connectivity, and high gate noise place strong pressure on circuit depth, two-qubit gate count, and hardware compatibility. In the longer term~\cite{fowler2012surface,forster2025quantum,zimboras2025myths}, fault-tolerant quantum computing will relax some hardware constraints but will not remove the need for efficient circuit synthesis, compilation, and task-adapted circuit design. Across both settings, the ability to construct compact and effective quantum circuits remains a key determinant of performance, resource cost, and practical utility.
\begin{figure}[h!]
    \centering
    \caption{\small \textbf{Overview of replay-buffer engineering for quantum optimization.}
(Left) \emph{Buffer engineering} improves experience reuse through replay design and sampling. (Middle) \emph{Amortized learning} reduces the cost of curriculum RL-based quantum architecture search by performing expensive quantum-classical updates only every $m$ steps.
(Right) \emph{Noise-aware transfer} warm-starts the RL-training in noisy environment by reusing trajectories collected in noiseless training.}    \includegraphics[width=0.9\linewidth]{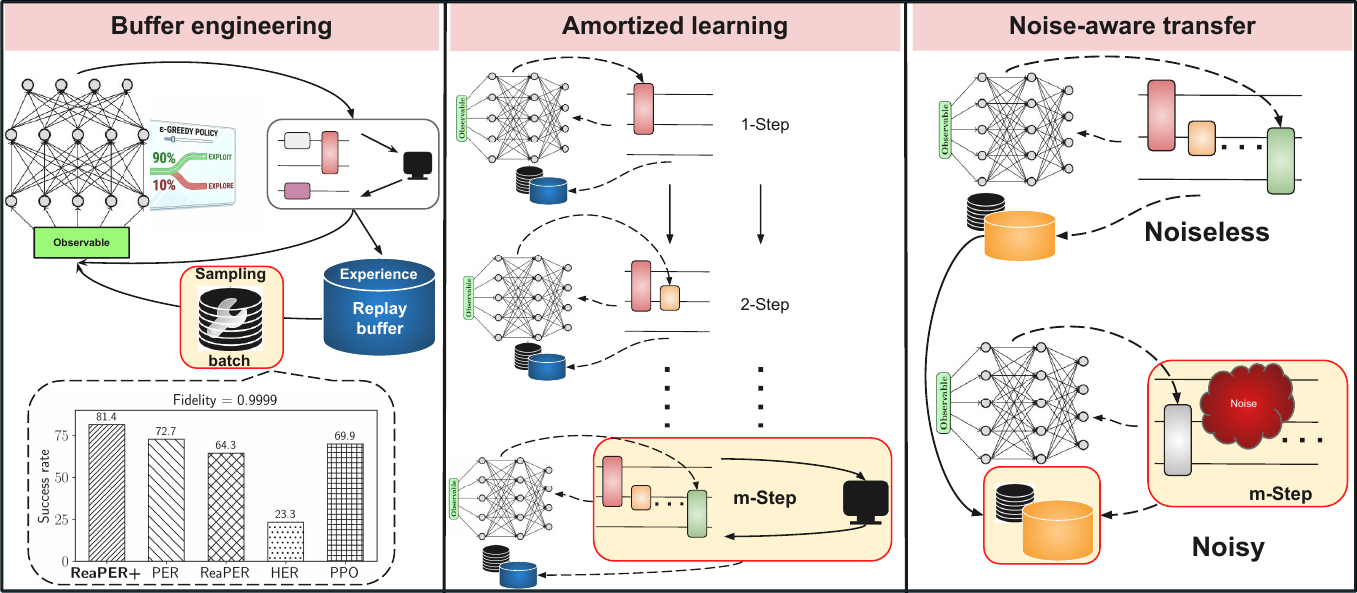}
    \label{fig:framework_ammortized_RL_transfer}
\end{figure}

Reinforcement learning (RL)~\cite{sutton1998reinforcement} has emerged as a promising framework for this problem because it treats circuit design as a sequential decision process, enabling agents to construct gate sequences or parameterized circuit architectures step by step~\cite{ostaszewski2021reinforcement, bukov2026reinforcement}. RL has been successfully applied to quantum compilation~\cite{moro2021quantum,wang2024quantum} and variational quantum circuit design~\cite{ostaszewski2021reinforcement, kundu2024reinforcement}. However, its practical use remains limited by a basic inefficiency: \emph{past experience is often poorly reused}. In most existing workflows~\cite{ostaszewski2021reinforcement,patel2024curriculum,sweke2021reinforcement,reuer2023realizing}, each new setting is treated as a fresh training problem, so trajectories 
collected in a noiseless simulator are discarded when the task is revisited under hardware noise which is an especially costly choice in quantum optimization~\cite{leblond2024realistic}, where the noiseless-to-noisy gap represents a shift between fundamentally distinct computational substrates that is physically motivated, practically unavoidable, and grows in severity with system size, making experience reuse both more challenging and more valuable than in classical transfer settings. 
In the standard variational framework~\cite{cerezo2021variational}, a parameterized quantum circuit $U(\boldsymbol{\theta},\mathcal{A})$ is defined by a discrete architecture $\mathcal{A}$ and 
continuous parameters $\boldsymbol{\theta}$, preparing a state 
$\ket{\psi(\boldsymbol{\theta},\mathcal{A})} = U(\boldsymbol{\theta},\mathcal{A})\ket{0}^{\otimes n}$, 
with cost $C(\boldsymbol{\theta},\mathcal{A}) = 
\bra{\psi(\boldsymbol{\theta},\mathcal{A})}H\ket{\psi(\boldsymbol{\theta},\mathcal{A})}$ estimated 
on the QPU and a classical optimizer updating $\boldsymbol{\theta}$ toward targets 
$d(U(\boldsymbol{\theta},\mathcal{A}),U_{\mathrm{tar}}) \le \varepsilon_{\mathrm{comp}}$ or 
$C(\boldsymbol{\theta},\mathcal{A}) - E_0 \le \varepsilon_{\mathrm{chem}}$. Each RL step may 
trigger a full quantum-classical evaluation, so wall-clock time grows rapidly with system size. 
While TensorRL-QAS~\cite{kundu2025tensorrl} narrows the search space via tensor network 
warm-starts, it still incurs substantial per-step cost and relies on standard replay buffers without explicit experience transfer; similarly, transfer in RL-optimized error-correcting codes~\cite{zen2025quantum} does not treat replay-buffer design as the primary transfer vehicle. As a result, the practicality of RL-based quantum optimization depends not only on the agent architecture, but critically on \emph{how experience is stored, prioritized, and transferred across the noiseless-to-noisy scenario} which is a direction that is not explored yet in the scope of quantum optimization.


We address these bottlenecks by treating the replay buffer as a primary
algorithmic lever for RL-based quantum optimization, motivated by evidence
from classical RL that buffer composition and sampling strongly affect
stability and sample efficiency~\cite{hester2018deep,zhou2025efficient,
lee2021offlinetoonline,10210487,song2025adaptivereplaybufferofflinetoonline}.
Figure~\ref{fig:framework_ammortized_RL_transfer} summarizes our framework,
combining annealed replay, amortized curriculum learning, and lightweight
noiseless-to-noisy transfer. Concretely, we make three contributions:

\begin{enumerate}[noitemsep,nolistsep]
\item \textbf{Annealed replay for quantum optimization.}
Under a fixed DQN~\cite{mnih2013playing}/DDQN~\cite{van2016deep} setup, we introduce \emph{ReaPER$+$}, an annealed replay rule transitioning from TD error prioritization~\cite{schaul2015prioritized} to reliability-aware replay~\cite{pleiss2026reliabilityadjusted}. It achieves up to $4\times$
greater sample efficiency over the strongest off-policy baselines (HER~\cite{andrychowicz2017hindsight}, PER~\cite{schaul2015prioritized}, and ReaPER~\cite{pleiss2026reliabilityadjusted}) and matches on-policy (PPO) solution quality at $32\times$ fewer interactions; a LunarLander-v3 validation (Appendix~\ref{app:lunarlander})
confirms it is not domain-specific. {Under a stated variance-alignment condition, ReaPER$+$ reduces the asymptotic noise floor versus PER (Theorem~\ref{thm:sample-complexity}, Appendix~\ref{appndx:reaper+}), a conditional, not unconditional, result. In Appendix~\ref{appndx:reaper_impl}, Algorithm~\ref{alg:reaper_plus_impl} gives the exact implementation, noting three deviations from this idealized form: a reward-based priority warm-start, an approximate online reliability update, and the importance-sampling
weighting used in the loss.}

\item \textbf{Amortized curriculum learning.}
\textit{OptCRLQAS}, an amortized variant of CRLQAS~\cite{patel2024curriculum},
reuses quantum-classical evaluations across multiple architectural edits,
cutting average wall-clock time per episode by $67.5\%$ ($3\times$ faster) on
12-qubit $\htwoo$ without degrading final energy error or gate count.

\item \textbf{Replay-based transfer to noise.}
A lightweight transfer scheme reuses noiseless trajectories to warm-start
noisy-setting learning, without network-weight transfer or long
$\epsilon$-greedy pretraining. On 6-qubit $\behtwo$ and 8-/12-qubit $\htwoo$
tasks, this buffer-only transfer cuts steps to chemical accuracy by up to
$85$-$90\%$ and final energy error by up to $90\%$ over from-scratch noisy
baselines, with the advantage growing with system size.
\end{enumerate}

\section{Related work}
\vspace{-8pt}
Quantum circuit optimization~\cite{abbas2024challenges} has been approached via adaptive ansatz construction~\cite{grimsley2019adaptive}, differentiable architecture search~\cite{zhang2022differentiable,wu2023quantumdarts}, evolutionary heuristics~\cite{rasconi2019innovative,sunkel2023ga4qco}, Bayesian optimization~\cite{nicoli2023physics}, Monte Carlo tree search~\cite{wang2023automated}, neural predictors~\cite{zhang2021neural}, and sampling-based strategies~\cite{du2022quantum}, demonstrating automatic circuit structure discovery but revealing persistent challenges
in scaling and noise robustness. Within this landscape, reinforcement learning has emerged as a flexible framework for both quantum compiling~\cite{moro2021quantum} and quantum architecture search~\cite{ostaszewski2021reinforcement,fosel2021quantum,kuo2021quantum}, with value-based and policy-gradient methods applied to variational ground state preparation~\cite{ostaszewski2021reinforcement,patel2024curriculum,kundu2025tensorrl}, entangled-state generation~\cite{kuo2021quantum,kundu2024kanqas}, and hardware-aware circuit design~\cite{kremer2024practical,kundu2026reinforcement}. 
However, existing RL-based frameworks treat each new setting as a fresh training instance, when a task moves from a noiseless simulator to noisy hardware, the agent is retrained from scratch
and accumulated experience is discarded, incurring substantial GPU and CPU cost per retraining cycle.

Experience replay is a central component of off-policy deep RL~\cite{mnih2013playing}, with uniform replay, hindsight experience replay~\cite{andrychowicz2017hindsight}, prioritized experience
replay~\cite{schaul2015prioritized}, and reliability-aware variants~\cite{pleiss2026reliabilityadjusted} providing increasingly structured
trade-offs between coverage, informativeness, and target reliability. In classical RL,
recent offline-to-online methods~\cite{zhou2025efficient,lee2021offlinetoonline,
10210487,song2025adaptivereplaybufferofflinetoonline} demonstrate that buffer composition and sampling rules are critical for stable, sample-efficient fine-tuning. In quantum optimization~\cite{patel2024curriculum,zen2025quantum}, however, replay buffers have largely been treated as fixed implementation choices rather than as a primary algorithmic design lever. This is the gap our work directly addresses by treating the replay buffer as the central object of design for quantum circuit optimization.

\section{Methods}\label{sec:methods}
\vspace{-8pt}
We introduce a replay-buffer engineering algorithm for quantum circuit optimization (Figure~\ref{fig:framework_ammortized_RL_transfer}) with three components: annealed replay (\textit{ReaPER+}), amortized curriculum learning (\textit{OptCRLQAS}), and lightweight noiseless-to-noisy buffer transfer. All experiments fix a common DQN/DDQN agent with identical state representation, action space, reward, and training protocol where only the replay mechanism is varied.

\paragraph{ReaPER$+$.} Among existing strategies, PER and ReaPER emerge as complementary baselines. PER (see Equation~\ref{eq:PER_sample}) prioritizes transitions
with large TD errors, providing aggressive early exploration but potentially amplifying noisy targets. ReaPER (see Equation~\ref{eq:ReaPER_sample}) discounts transitions whose downstream TD errors indicate unreliable targets, better matching the long-horizon structure of episodic quantum optimization but
converging more slowly in the early phase. This complementary behavior motivates a hybrid that transitions from PER-like exploration to ReaPER-like refinement over the course of training.

To exploit the best of both regimes, we introduce \emph{ReaPER+}, a 
replay strategy that transitions smoothly from PER-like prioritization 
to ReaPER-like prioritization over the course of training. Specifically, 
at training step $\tau$ we define
\begin{equation}
    \Psi_t^{(+,\tau)}
    =
    \mathcal{R}_t^{\omega_{\tau}} (\delta_t^{+})^{\alpha},
    \qquad
    \mu_t^{(+,\tau)}
    =
    \frac{\Psi_t^{(+,\tau)}}{\sum_i \Psi_i^{(+,\tau)}},
\end{equation}
where $\mathcal{R}_t$ is the reliability score at step $t$ (described more elaborately in Appendix~\ref{appndx:preliminaries}), the annealing exponent $\omega_{\tau} \in [0,1]$ is non-decreasing 
in $\tau$. $\omega_{\tau}=0$ recovers PER and $\omega_{\tau}=1$ recovers 
ReaPER exactly. In practice, we use the linear schedule
\begin{equation}
    \omega_{\tau}
    =
    \omega_{\min}
    +
    (\omega_{\max}-\omega_{\min})
    \min\!\left(\frac{\tau}{T_{\mathrm{ann}}},\,1\right),
    \label{ref:anneal_schedule}
\end{equation}
which is monotone and controlled by a single interpretable timescale 
$T_{\mathrm{ann}}$, avoiding the rapid early transitions of cosine or 
exponential schedules when reliability estimates are still poorly 
calibrated. We set $\omega_{\min}=0.1>0$ so that a residual TD error 
signal is retained at initialization, where $\mathcal{R}_t$ carries 
little meaning under a random $Q$-function, and $\omega_{\max}=0.7<1$ 
to prevent premature over-commitment to reliability scores before $Q$-function convergence values $\omega_{max}\ge0.9$ produced slower convergence in both settings; see Appendix~\ref{appndx:sensitivity_1q}. $T_{\mathrm{ann}}$ is set to 
half the total training budget ($5\times10^{5}$ for compilation, 
$5\times10^{4}$ for LunarLander-v3; Table~\ref{tab:hyperparams}), so 
the transition completes by mid-training when value estimates are 
sufficiently stable. Early in training sampling is thus driven by TD 
error; later, the influence of $\mathcal{R}_t$ grows, biasing replay toward 
transitions that are both informative and reliable. ReaPER+ therefore 
preserves the sample-efficiency advantages of PER at the beginning of 
learning while inheriting the stability of ReaPER once value estimates 
mature, a design choice directly motivated by the empirical finding 
that PER excels early while ReaPER is preferable once estimates 
stabilize {this is confirmed by our systematic benchmark of uniform replay, HER~\cite{andrychowicz2017hindsight}, PER~\cite{schaul2015prioritized}, and ReaPER~\cite{pleiss2026reliabilityadjusted} under a fixed agent across all tasks. Under a stated variance-alignment condition on the reliability score, ReaPER$+$ reduces the asymptotic noise floor compared to PER in Theorem~\ref{thm:sample-complexity} (Appendix~\ref{appndx:reaper+}); we present this as a conditional theoretical result rather than an unconditional guarantee.}

\paragraph{Lightweight buffer transfer.} In addition to replay-buffer engineering, we study a \textit{replay 
buffer transfer} scheme for noise-aware quantum optimization. The idea is to first collect trajectories in a source environment, typically the 
noiseless version of a given compiling or QAS task, and then use the 
resulting replay memory to initialize training in the corresponding noisy target environment. Let $\mathcal{B}_{\mathrm{src}}$ denote the replay buffer obtained after source training and $\mathcal{B}_{\mathrm{tgt}}^{(0)}$ the initial replay buffer of the target task. We initialize the target buffer directly as
\begin{equation}
    \mathcal{B}^{(0)}_{\mathrm{tgt}} \leftarrow \mathcal{B}_{\mathrm{src}},
    \label{eq:transfer}
\end{equation}
that is, all stored transitions $(S_t, A_t, R_t, S_{t+1}, d_t)$ are 
copied from the source buffer to the target buffer without modification, 
relabeling, or filtering. This is valid because the noiseless and noisy 
environments share identical state and action spaces: the RL-state and the gate action set $\mathcal{A}$ are unchanged by the introduction of noise, which only affects the transition dynamics and reward statistics. Buffer-only transfer of this kind, where source experience warm-starts a target replay buffer without sharing network weights, has recently proved effective in classical deep RL. Zhou et al.~\cite{zhou2025efficient} show that seeding the online replay buffer with offline transitions suffices for sample-efficient fine-tuning without retaining the offline dataset or transferring network parameters. Lee et al.~\cite{lee2021offlinetoonline} study balanced mixing of offline and online replay in the offline-to-online setting, finding that the buffer composition critically determines early learning speed. Our scheme follows the same instance-transfer paradigm~\cite{you2022cross}: 
experience is transferred at the trajectory level rather than the parameter 
level. We deliberately decouple buffer and weight transfer, as network weights trained under noiseless dynamics may introduce systematic bias when the target environment includes depolarizing noise which is a risk that does not arise with buffer-only transfer, where source transitions are corrected online as the agent accumulates noisy target experience.

Consequently, trajectories that are informative in the noiseless setting can still provide a strong initialization for learning under noise by improving early buffer coverage and accelerating the discovery of high-quality circuits. In Appendix~\ref{appndx:buffer-transfer} we provide a formal argument for why replay-buffer transfer is natural in quantum optimization settings under the assumption that source and target tasks share approximately similar state and action spaces.

\paragraph{OptCRLQAS.} Finally, to scale while operating with CRLQAS we realized the time per 
episode scales rapidly with qubit count, and training the RL-agent with 
CRLQAS beyond 10-qubit can take a substantial amount of GPU compute. To 
address this issue we introduce \emph{OptCRLQAS}, an efficiency-oriented 
variant of curriculum reinforcement learning-based quantum architecture 
search. In standard CRLQAS, each environment step triggers a full 
quantum-classical evaluation of the current circuit, including variational parameter optimization and cost-function estimation, so for an episode of length $T$, the total evaluation cost scales as $T \cdot C_{\mathrm{eval}}$, where $C_{\mathrm{eval}}$ denotes the cost of a single quantum-classical call. OptCRLQAS reduces this by \textit{amortizing} each evaluation over $m$ consecutive architectural edits: rather than invoking a new variational optimization at every step, the agent accumulates $m$ local gate modifications before triggering a single evaluation, spreading the cost $C_{\mathrm{eval}}$ 
across $m$ steps. Formally, let $u_\tau \in \{0,1\}$ denote the update 
indicator,
\begin{equation}
    u_\tau =
    \begin{cases}
        1, & \text{if } \tau \bmod m = 0 \text{ or the episode terminates},\\
        0, & \text{otherwise},
    \end{cases}
\end{equation}
so that a full architecture evaluation is performed only when $u_\tau = 1$. Hence, for an episode of length $T$, the number of expensive 
quantum-classical evaluations is reduced from $T$ in CRLQAS to 
approximately $\lceil T/m \rceil$ in OptCRLQAS, yielding an expected 
reduction by a factor of about $m$ when episode lengths are sufficiently 
large.

Beyond the computational saving, accumulating $m$ edits before evaluating also improves the learning signal: single-gate modifications are often too small to produce distinguishable rewards when variational parameters $\boldsymbol{\theta}$ can compensate local changes, leading to weakly separated TD targets and slow value propagation. Judging a block of $m$ edits jointly increases reward contrast and allows useful gate combinations to be credited as a unit. A more detailed discussion of both effects is provided in Section~\ref{sec:qas}. Throughout the paper, \emph{OptCRLQAS} refers to $m=10$ unless stated otherwise; a quality-vs-runtime sweep over $m\in\{3,5,7,10\}$ in Appendix~\ref{appndx:m_sweep} confirms this efficiency gain holds across the
full range without systematically degrading solution quality.
\begin{figure}[b!]
  \centering
  \footnotesize
  \caption{\small \textbf{1-qubit compiling of Haar-random target unitaries with
$\texttt{RX},\texttt{RY},\texttt{RZ}(\pm\pi/128)$ gates.} (Left) Success probability and mean fidelity at tolerances $0.99$, $0.999$, and $0.9999$, where ReaPER$+$ performs best overall. HER~\cite{andrychowicz2017hindsight} uses uniform (non-prioritized) sampling over its goal-relabeled buffer and serves as a uniform-replay reference point. (Right) mean circuit length with $\pm 1\sigma$ error bars versus tolerance; although all methods require deeper circuits at higher accuracy and exhibit a similar growth rate with tightening tolerance, ReaPER$+$ maintains a consistently lower circuit-length offset, giving the best accuracy-length tradeoff across all tolerance levels.}
  \begin{minipage}{0.05\linewidth}
    \centering
    \begin{tabular}{l l c c}
      \toprule
      \textbf{Tol} & \textbf{Method} & \textbf{Success} (\%) $\uparrow$ & \textbf{Avg. fidelity} $\uparrow$ \\
      \midrule
      0.99 & \textbf{ReaPER+ (Ours)} & \textbf{89.30} & \textbf{0.94} \\
           & PER~\cite{schaul2015prioritized}           & 85.81 & 0.94 \\
           & ReaPER~\cite{pleiss2026reliabilityadjusted}        & 85.15 & 0.93 \\
           & HER~\cite{andrychowicz2017hindsight}        & 76.49 & 0.89 \\
           & PPO~\cite{schulman2017proximal}           & 75.40 & 0.88 \\
      \midrule
      0.999 & \textbf{ReaPER+ (Ours)} & \textbf{85.30} & \textbf{0.95} \\
            & PER           & 82.35 & 0.94 \\
            & ReaPER        & 80.97 & 0.93 \\
            & HER        & 74.44 & 0.90 \\
            & PPO           & 71.60 & 0.87 \\
      \midrule
      0.9999 & \textbf{ReaPER+ (Ours)} & \textbf{81.40} & \textbf{0.95} \\
             & PER           & 72.66 & 0.94 \\
             & ReaPER        & 64.35 & 0.93 \\
             & HER        & 63.30 & 0.90 \\
             & PPO           & 69.90 & 0.87 \\
      \bottomrule
    \end{tabular}
  \end{minipage}%
  \hfill
  \begin{minipage}{0.4\linewidth}
    \centering
    \includegraphics[width=\linewidth]{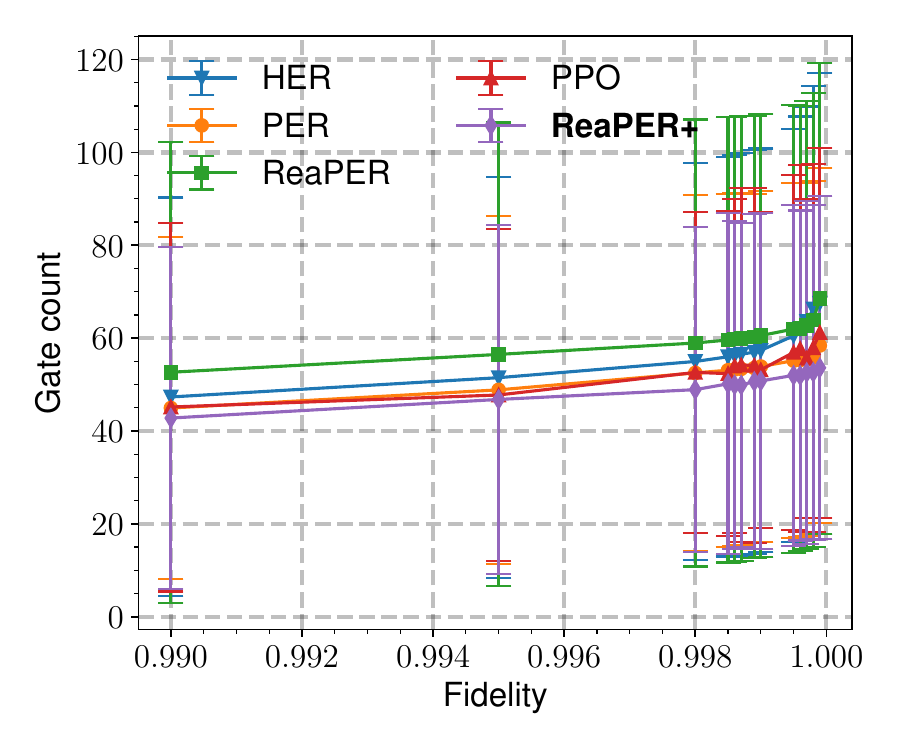}
  \end{minipage}
  \label{fig:haar1q-summary}
\end{figure}

\section{Results}\label{sec:results}
\vspace{-8pt}
We evaluate replay-buffer design in quantum compilation, where the agent's task is to synthesize a target unitary encoded in an RL-state of size $2\times2^N$ by sequentially appending gates from a fixed action space defined in Section~\ref{sec:state_action_compiling}. In every training episode the environment resets to the $\ket{0}$ state and a fresh Haar-random 1-qubit target unitary is sampled; the agent acts until the target fidelity tolerance is met or the maximum episode length $L$ is reached. All agents are trained for $5\times10^{4}$ episodes and evaluated on $10^{5}$ independently sampled Haar-random target unitaries. Full hyperparameter details are given in Appendix~\ref{appndx:hyperparams}. Moreover, the compute details are provided in Appendix~\ref{appndx:cpu_gpu_details}.

\paragraph{Experimental setup and reproducibility.} All experiments follow a common pipeline: an RL agent (DQN~\cite{mnih2013playing}/DDQN~\cite{van2016deep}, Section~\ref{sec:methods}) proposes a gate action, the resulting circuit is evaluated by a quantum simulator. The parameter of quantum circuit optimized using \texttt{COBYLA}~\cite{powell1994direct} and  then quantum simulator does cost function estimation. Finally the resulting cost/fidelity signal is returned to the agent as a reward. Quantum compiling
experiments (Section~\ref{subsec:quantum_compiling}) use \texttt{Qiskit}~\cite{javadi2024quantum}, while quantum
architecture search experiments (Section~\ref{sec:qas}) use \texttt{Qulacs}~\cite{suzuki2021qulacs} for faster state vector simulation at larger qubit counts. Noise is injected at the simulator level via depolarizing and amplitude damping channels (Section~\ref{subsec:buffer_transfer}). All experiments were run on local and HPC machines, with per-experiment runtimes not exceeding 48 hours; hardware configurations and wall-clock time for every experiment category are summarized in Appendix~\ref{appndx:cpu_gpu_details}. Complete hyperparameters are listed in Appendix~\ref{appndx:hyperparams}, molecule geometries and Hamiltonians in Appendix~\ref{tab:molecule_geometry}, and the full noise-modeling stack in Section~\ref{subsec:buffer_transfer} and Appendix~\ref{appndx:buffer-transfer}. Our complete training code, environment wrappers, and evaluation scripts will be publicly released to
enable full reproduction of every result in this paper.

\subsection{Quantum compiling}
\label{subsec:quantum_compiling}
\vspace{-4pt}
\paragraph{1-qubit compiling.}
We evaluate two settings: (i)~the small-rotation basis
$\texttt{RX},\texttt{RY},\texttt{RZ}(\pm\pi/128)$ and (ii)~the discrete HRC gate
set~\cite{harrow2002efficient}, following Ref.~\cite{moro2021quantum}.
Figure~\ref{fig:haar1q-summary} reports success probability and fidelity over
tolerances $0.99$-$0.9999$ (40 seeds). ReaPER$+$ achieves the highest success at
every tolerance ($89.30\%$, $85.30\%$, $81.40\%$), outperforming PER ($85.81\%$,
$82.35\%$, $72.66\%$) and fixed ReaPER ($85.15\%$, $80.97\%$, $64.35\%$), while
showing the slowest circuit-length growth, yielding the best success-fidelity-length
tradeoff overall. Unlike PER, ReaPER, and ReaPER$+$, which differ in how
transitions are \emph{prioritized}, HER instead samples uniformly from a
goal-relabeled buffer, so it serves here as a uniform-replay reference point
rather than a fourth prioritized variant. To assess the reliability of this
circuit-length advantage, we additionally conduct a sample significance analysis in Appendix~\ref{appndx:1q_significance}. While the resulting 95\% confidence intervals are wide due to the small sample size, the analysis confirms a consistent downward shift in gate count for ReaPER$+$ relative to PER and fixed ReaPER across all three tolerances. For the HRC basis, Table~\ref{tab:hrc_1q_haar} shows that ReaPER$+$ reaches $100\%$
success with mean fidelity $0.995$ and shortest circuits ($14.30\pm7.89$ gates) at $1.56\times10^6$ steps, $\sim\!24\%$ faster than fixed ReaPER, $\sim26\%$ faster than PER, and $\sim\!72\%$ faster than our tuned HER baseline ($k{=}5$ relabelings,
100-episode warm-up), which despite being stronger than the original~\cite{moro2021quantum}
still plateaus at $95\%$ success. All subsequent ``HER'' references denote this
tuned implementation.
\vspace{-6pt}
\paragraph{2-qubit compiling.}
Using target unitaries sampled via Algorithm~\ref{alg:2qubit_target}, we test whether
agents can approximate a $\texttt{ZZ}(\pi)$ gate at fidelity threshold $0.9914$,
matching the benchmark of Ref.~\cite{moro2021quantum}.
Table~\ref{tab:2q_haar} shows ReaPER$+$ reaches fidelity $0.9920$ in only
$2.5\times10^4$ episodes, a $4\times$ reduction over fixed ReaPER, PER, and HER
(all at $10^5$ episodes), and a $\mathbf{32\times}$ reduction over
PPO~\cite{moro2021quantum} ($8\times10^5$ episodes). Although fixed ReaPER attains a
marginally higher best fidelity ($0.9931$) given its longer budget, ReaPER$+$ reaches
a comparable fidelity regime far more efficiently, the operationally relevant metric
when quantum-classical simulation time is limited.
\begin{table}[h!]
\centering
\footnotesize
\caption{\small \textbf{ReaPER+ outperforms all baselines on $ZZ(\pi)$ approximation
using 2-qubit gates.} ReaPER$+$ matches or exceeds fidelity at $32\times$ fewer
episodes than PPO~\cite{moro2021quantum}. "NA" means the with HER buffer we could not achieve a fidelity below the pre-defined fidelity threshold $0.9914$ with fixed computational budget.}
\label{tab:2q_haar}
\begin{tabular}{lrrr}
\toprule
\textbf{Method} & \textbf{Episodes} $\downarrow$ & \textbf{Min.\ gates} $\downarrow$
& \textbf{Best fidelity} $\uparrow$ \\
\midrule
\textbf{ReaPER+ (Ours)}        & $\mathbf{2.5\times10^{4}}$ & $\mathbf{123}$ & $0.9920$ \\
ReaPER~\cite{pleiss2026reliabilityadjusted} & $10^{5}$ & $126$ & $\mathbf{0.9931}$ \\
PPO (Moro et al.~\cite{moro2021quantum})   & $8\times10^{5}$ & $122$ & $0.9914$ \\
PER~\cite{schaul2015prioritized}           & $10^{5}$ & $127$ & $0.9918$ \\
HER~\cite{andrychowicz2017hindsight}       & $10^{5}$ & NA    & $<0.9914$ \\
\bottomrule
\end{tabular}
\end{table}

\vspace{-6pt}
\subsection{Quantum architecture search}
\label{sec:qas}
\vspace{-8pt}
We evaluate replay buffers in QAS, where the agent seeks a parameterized circuit
that prepares the ground state of a target molecular Hamiltonian. Following
CRLQAS~\cite{patel2024curriculum,ostaszewski2021reinforcement}, we use the gate set
in Equation~\ref{eq:qas_gateset} and reward in Equation~\ref{eq:reward_qas}, benchmarking
molecular ground state preparation tasks of increasing scale (molecule geometries and
hyperparameters in Appendices~\ref{tab:molecule_geometry} and 
\ref{appndx:agent_env_hyperparameter_QAS}, respectively).

\begin{table}[t]
\footnotesize
\centering
\caption{\small\textbf{OptCRLQAS + ReaPER$+$ vs.\ non-RL baselines.} We report mean $\pm$ standard deviation over 5 random seeds for our method. Our method achieves the lowest energy error on 2 of 3 systems while using competitive or fewer gates and CNOTs. The 5-qubit Heisenberg model is described in Appendix~\ref{appndx:heisenberg}. For 5-Heisenberg we utilize an action space consisting of $\{\texttt{XX}(\theta_{xx}), \texttt{YY}(\theta_{yy}), \texttt{ZZ}(\theta_{zz}), \texttt{RX}(\theta_{x}), \texttt{RY}(\theta_{y}), \texttt{RZ}(\theta_{z})\}$ gates. Baseline results were not re-run (NR) as per-seed statistics/code were unavailable, except SA-QAS, which we re-ran ourselves (R).}
\label{tab:6q_comparison}
\begin{tabular}{lcccc}
\toprule
Problem & Method & Final error (Ha) & Total gates & CNOT \\
\midrule
5-Heisenberg
  & \textbf{OptCRLQAS + ReaPER$+$ (ours)} & $\mathbf{(5.9\pm0.7)\times10^{-4}}$ & $41.0\pm7.5$ & NA \\
  & DQAS~\cite{zhang2022differentiable} (NR)    & $1.1\times10^{-1}$          & 35  & NA \\
  & GQAS~\cite{he2024gradient} (NR)             & $7.1\times10^{-4}$          & 35  & NA \\
  & TF-QAS~\cite{he2024training} (NR)           & $1.2\times10^{-3}$          & 35  & NA \\
\midrule
6-$\behtwo$
  & \textbf{OptCRLQAS + ReaPER$+$ (ours)} & $\mathbf{(6.2\pm3.3)\times10^{-5}}$ & $\mathbf{52.8\pm14.4}$ & $\mathbf{24\pm6.1}$ \\
  & TF-QAS~\cite{he2024training} (NR)           & $1.8\times10^{-3}$          & 57          & NA \\
  & SA-QAS~\cite{lu2023qas} (R)                & $5.6\times10^{-3}$          & 73 & 45 \\
\midrule
8-$\htwoo$
  & \textbf{OptCRLQAS + ReaPER$+$ (ours)} & $(2.2\pm1.4)\times10^{-4}$ & $\mathbf{89.8\pm8.6}$ & $\mathbf{61.2\pm12.6}$ \\
  & quantumDARTS~\cite{wu2023quantumdarts} (NR) & $\mathbf{1.7\times10^{-4}}$          & 219          & 68 \\
  & SA-QAS~\cite{lu2023qas} (R)                & $2.6\times10^{-3}$          & 95           & 69 \\
\bottomrule
\end{tabular}
\end{table}

\begin{figure}[t!]
    \centering
    \caption{\small \textbf{Replay-buffer design controls circuit compactness in QAS.} For 6-\behtwo\ and 8-\htwoo, ReaPER$+$ variants yield the lowest total, \texttt{CNOT}, and rotation gate counts compared to PER and uniform replay (mean $\pm$ std over seeds). $\omega{=}0$ recovers PER; $\omega{=}1$ gives fully reliability-adjusted replay. Note that the error bars are 1-sigma.}
    \includegraphics[width=\linewidth]{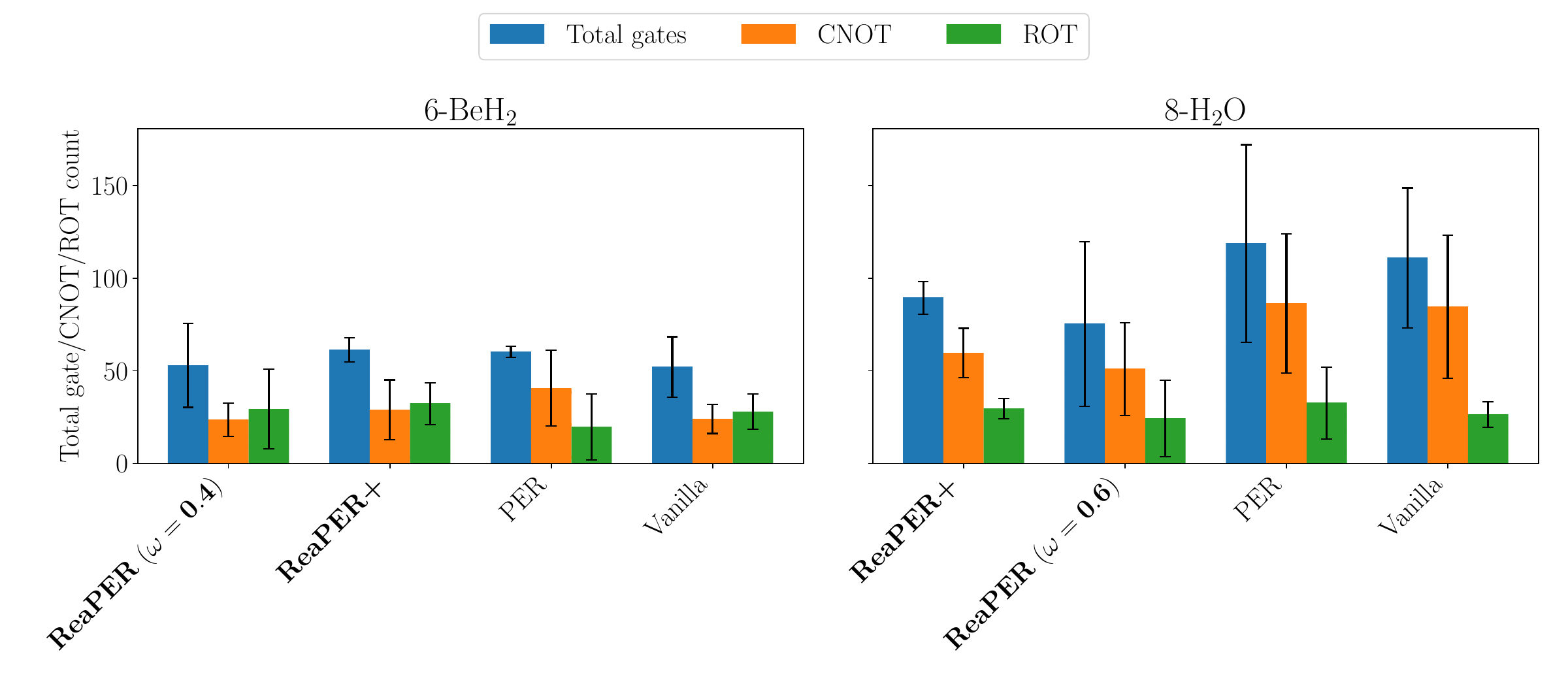}
    \label{fig:6q_8q_buffer}
\end{figure}

Figure~\ref{fig:6q_8q_buffer} compares replay strategies on the smaller-scale systems.
ReaPER$+$ achieves the lowest energy error among prioritized methods with competitive
circuit compactness; fixed ReaPER ($\omega{=}0.4$ for \behtwo, $\omega{=}0.6$ for
\htwoo) produces the most compact circuits, reflecting the longer-horizon credit
assignment at larger scale (full $\omega$ sensitivity in
Appendix~\ref{appndx:omega_selection}). Uniform replay yields superficially shorter circuits but at substantially higher energy error, indicating early trapping in local minima. As shown in Table~\ref{tab:6q_comparison}, OptCRLQAS with ReaPER$+$ (denoted as ``OptCRLQAS + ReaPER$+$") achieves the lowest energy error on the 5- and 6-qubit QAS problems, and is competitive with quantumDARTS~\cite{wu2023quantumdarts} on the 8-qubit problem, outperforming non-RL baselines such as DQAS~\cite{zhang2022differentiable}, GQAS~\cite{he2024gradient}, TF-QAS~\cite{he2024training}, SA-QAS~\cite{lu2023qas}, and quantumDARTS~\cite{wu2023quantumdarts} while using competitive or fewer gates and CNOTs overall. Results for our method are reported as mean $\pm$ standard deviation over 5 random seeds. For the 5-qubit problem, we target the ground state energy of the Heisenberg model (Hamiltonian given in Appendix~\ref{appndx:heisenberg}). Since this problem is well-suited to the gateset $\{\texttt{RXX}(\theta_{xx}),\, \texttt{RYY}(\theta_{yy}),\,
\texttt{RZZ}(\theta_{zz}),\, \texttt{RX}(\theta_{x}),\,
\texttt{RY}(\theta_{y}),\, \texttt{RZ}(\theta_{z})\}$, we employ an alternative RL state encoding described in Appendix~\ref{appndx:state_encoding}.

At 12-qubit scale, standard CRLQAS is prohibitively expensive since every step
triggers a full variational optimization. OptCRLQAS amortizes this cost over $m$ steps, reducing evaluations per episode from $T$ to $\lceil T/m \rceil$ and cutting average wall-clock time by $67.5\%$ (${\sim}3\times$) without degrading solution quality (Figure~\ref{fig:h2o_performance_combined}). {A quality-vs-runtime sweep over the amortization interval $m$ in Appendix~\ref{appndx:m_sweep} confirms this efficiency gain holds across $m\in\{3,5,7,10\}$ without systematically degrading solution quality}. Batching $m$ edits also yields a more separable learning signal by crediting meaningful architectural blocks rather than nearly indistinguishable single-step edits.

\begin{figure}[h!]
    \centering
    \caption{\small \textbf{Efficiency and performance on 12-qubit H$_2$O.}
    \textbf{(Left)} OptCRLQAS reduces wall-clock time per episode by $67.5\%$ over CRLQAS~\cite{patel2024curriculum}. \textbf{(Right)} Results are averaged over multiple seeds (mean $\pm$ std.\ dev.). While fixed ReaPER attains the lowest mean energy error and fewest steps, ReaPER$+$ achieves a competitive energy error while using substantially fewer gates and CNOTs than all other baselines, resolving the single-seed anomaly reported in the original
    submission.}
    \begin{minipage}[c]{0.3\linewidth}
        \centering
        \includegraphics[width=\linewidth]{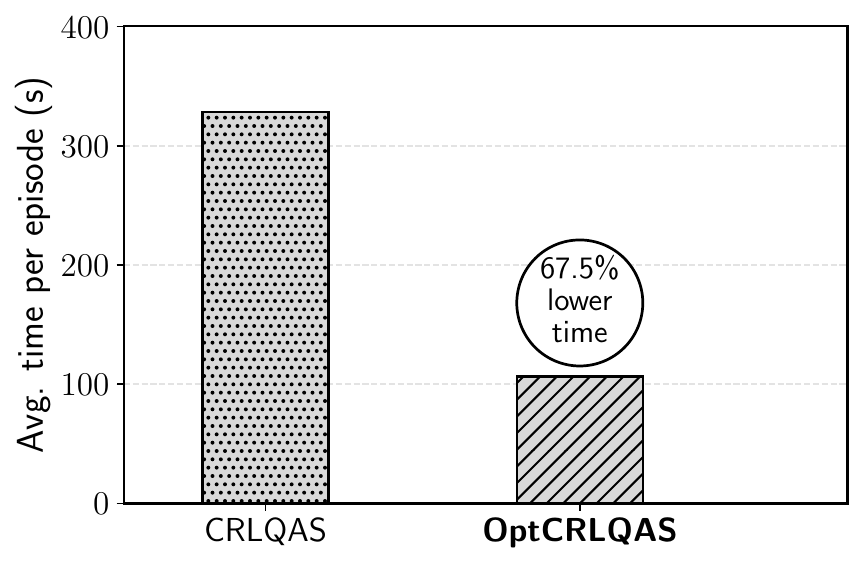}
    \end{minipage}\hfill
    \begin{minipage}[c]{0.7\linewidth}
        \centering\tiny
        \begin{tabular}{lcccc}
            \toprule
            \textbf{Method} & \textbf{Error} $\downarrow$ & \textbf{Gates} $\downarrow$
            & \textbf{\texttt{CNOT}} $\downarrow$ & \textbf{Steps} $\downarrow$ \\
            \midrule
            ReaPER      & $\mathbf{(2.00\pm0.26)\times10^{-2}}$ & $221\pm57$          & $125\pm29$          & $\mathbf{(2.5\pm2.3)\times10^{4}}$ \\
            \textbf{ReaPER$+$} & $(2.13\pm0.21)\times10^{-2}$          & $\mathbf{128\pm29}$ & $\mathbf{75\pm35}$  & $(3.7\pm2.0)\times10^{4}$ \\
            PER         & $(2.43\pm0.12)\times10^{-2}$          & $165\pm45$          & $95\pm28$           & $(5.7\pm3.2)\times10^{4}$ \\
            Vanilla     & $(2.45\pm0.20)\times10^{-2}$          & $151\pm35$          & $91\pm41$           & $(3.2\pm5.2)\times10^{4}$ \\
            \bottomrule
        \end{tabular}
    \end{minipage}
    \label{fig:h2o_performance_combined}
\end{figure}

With OptCRLQAS enabling 12-qubit training, ReaPER attains the lowest mean energy
error $(2.00\pm0.26)\times10^{-2}$ Ha in the fewest steps
($(2.5\pm2.3)\times10^{4}$), averaged over multiple seeds. ReaPER$+$ reaches a competitive energy error ($(2.13\pm0.21)\times10^{-2}$ Ha) while using
substantially fewer gates ($128\pm29$ vs.$221\pm57$) and CNOTs ($75\pm35$ vs. $125\pm29$) than fixed ReaPER, whereas PER reaches a similar error floor at roughly $2.3\times$ the step cost and vanilla replay converges to the worst solution among the four methods. We further validate that the $T_{ann}=T_{total}/2$ scaling rule remains reliable at 12-qubit scale in a targeted ablation in Appendix~\ref{appndx:sensitivity_12q}, which confirms that linking $T_{ann}$ to the training horizon minimizes total gate count without sacrificing energy accuracy. Appendix~\ref{appndx:optcrlqas_vs_crlqas} further shows OptCRLQAS can reduce quantum simulation time by up to $89\%$ and classical optimization time by up to $85\%$ in matched comparisons.

{\paragraph{Scalability analysis.} Beyond the 12-qubit ceiling of curriculum RL, combining ReaPER$+$ and amortized evaluation with TensorRL-QAS~\cite{kundu2025tensorrl} extends scaling to 20 qubits: with a smaller amortization step ($m{=}2$ at 15/17 qubits, $m{=}4$ at 20), wall-clock time per episode falls from $120.7$s at 12 qubits to $11.6$s at 15 qubits and only $100.4$s at 20, while the agent adds just $20$--$40$ compact edits atop the fixed MPS circuit. We report this as a structural proof-of-concept rather than a chemically accurate result, since errors here remain an order of magnitude above our chemical-accuracy threshold; full details in Appendix~\ref{appndx:tensorrl_scaling}.}
\vspace{-6pt}
\subsection{Noise-aware learning through buffer transfer}
\label{subsec:buffer_transfer}
\vspace{-8pt}
We evaluate the transfer scheme of Section~\ref{sec:methods}, where a buffer
$\mathcal{B}_{\mathrm{src}}$ collected in a noiseless source environment
initializes training in a related noisy target. This is, to our knowledge,
the first demonstration that a lightweight, weight-free replay buffer alone
transfers noiseless experience to realistic depolarizing noise settings, with
advantages growing with system size up to 12 qubits. All experiments use
OptCRLQAS with uniform replay to isolate transfer's effect. We study
(i)~\textit{noiseless-to-noiseless} and (ii)~\textit{noiseless-to-noisy}
transfer (depolarizing noise, strengths $p_1$/$p_2$) on molecular
ground state benchmarks of increasing scale. The source buffer comes from a
vanilla agent trained for a fixed 12 GPU-hour budget; to exploit the warm
start, we reduce initial exploration from $\epsilon{=}1.0$ to $0.55$ and
tighten the curriculum. We quantify transfer via the weighted score
\begin{equation}
S = w_1\,\Delta\text{steps} + w_2\,\Delta\text{ROT} + w_3\,\Delta\texttt{CNOT} +
w_4\,\Delta\text{err}.
\label{eq:score_transfer}
\end{equation}
Weighted scores are commonly used in benchmark studies~\cite{ikhtiarudin2025benchrl}.
In Equation~\ref{eq:score_transfer} weights $0.4, 0.1, 0.2, 0.3$, measures relative improvement over the no-transfer baseline. 

\begin{figure}[h!]
    \centering
    \caption{\small \textbf{Weighted transfer matrix for \behtwo\ under noiseless and noisy
    transfer.} Buffer transfer reduces steps to chemical accuracy by $47$-$58\%$ and
    improves final energy by up to $90.2\%$, yielding composite scores of
    $19.2$-$35.8\%$; the strongest score ($35.8\%$) comes from the largest
    energy improvement at $p_2{=}0.001$.}
    \includegraphics[width=0.9\linewidth]{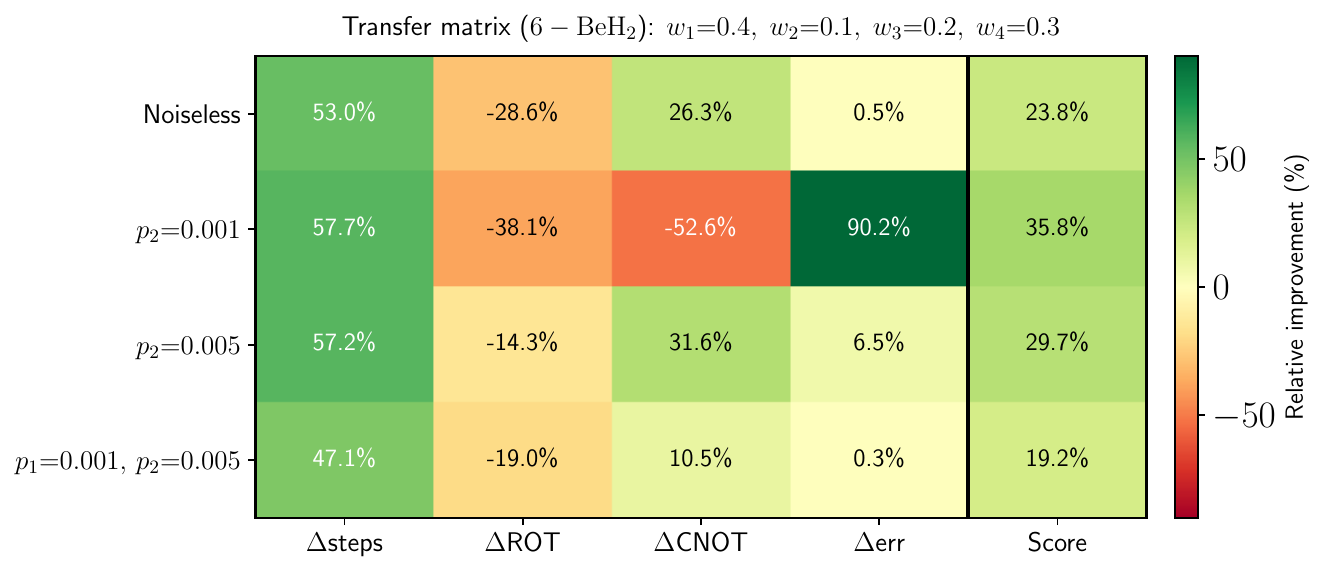}
    \label{fig:6q_buffer_transfer_matrix}
\end{figure}

\begin{figure}[h!]
    \centering
    \caption{\small \textbf{Weighted transfer matrix for $\htwoo$ under noiseless and noisy
    transfer.} Step reductions range from $49.8\%$ to $84.8\%$, with energy
    improvements up to $46.7\%$ under combined noise ($p_1{=}0.001$,
    $p_2{=}0.005$), yielding the highest score of $28.7\%$.}
    \includegraphics[width=0.9\linewidth]{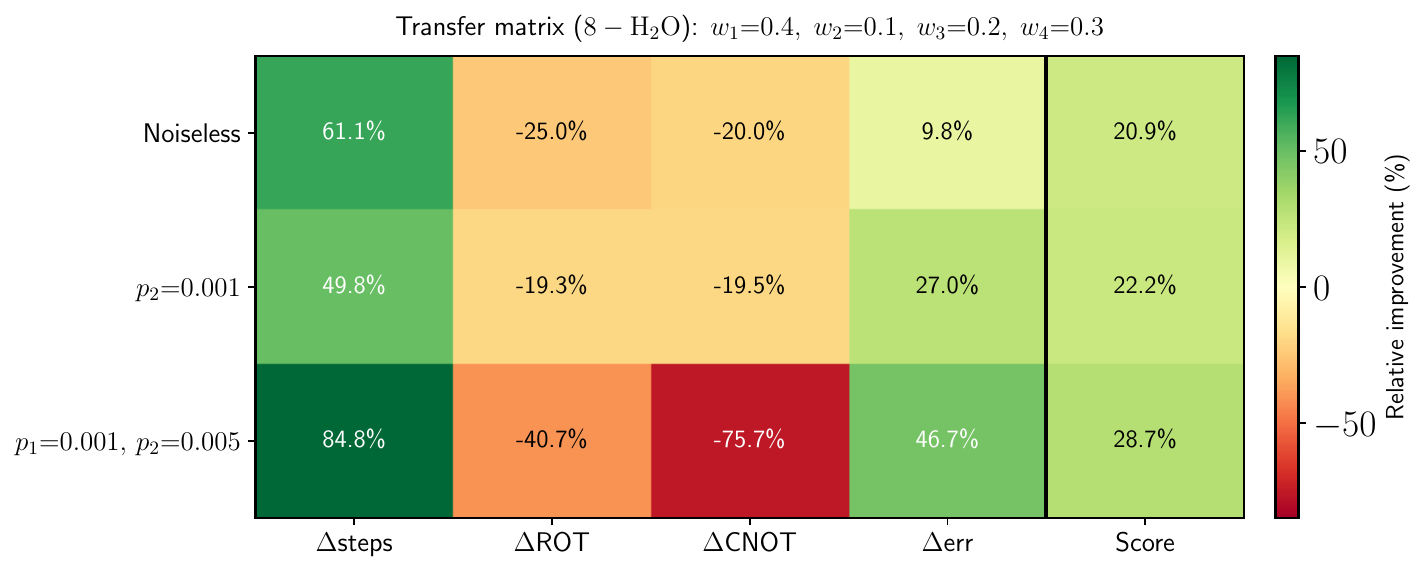}
    \label{fig:8q_buffer_transfer_matrix}
\end{figure}

\begin{figure}[h!]
    \centering
    \caption{\small \textbf{Weighted transfer matrix for 12-qubit $\htwoo$ under noiseless
    and noisy transfer.} Transfer reduces steps to match non-transfer accuracy
    by $88.2\%$ and CNOT count by $57.6\%$, under combined depolarizing noise
    ($p_1{=}0.001$, $p_2{=}0.005$).}
    \includegraphics[width=0.9\linewidth]{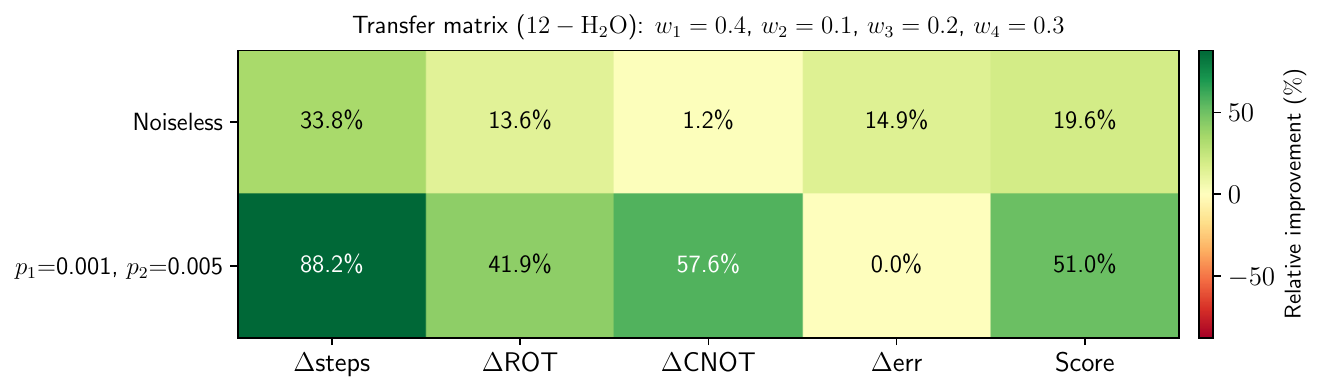}
    \label{fig:12q_buffer_transfer_matrix}
\end{figure}

Across all systems and noise settings, buffer transfer consistently reduces
steps to chemical accuracy and improves final energy error. At the smallest
scale (Figure~\ref{fig:6q_buffer_transfer_matrix}), step reductions of
$47$-$58\%$ and energy gains up to $90.2\%$ yield composite scores of
$19.2$-$35.8\%$. At 8 qubits (Figure~\ref{fig:8q_buffer_transfer_matrix}),
step reductions reach $84.8\%$ under combined noise. Crucially, at 12 qubits
(Figure~\ref{fig:12q_buffer_transfer_matrix}), \textit{the advantage
strengthens with system size}: transfer reduces steps by $88.2\%$ (without
error improvement), achieving the highest composite score of $51.0\%$ across
all benchmarks. Lightweight buffer transfer thus becomes \textit{more}
beneficial as the system grows, with \textit{no architectural changes, no
reward relabeling, and no network weight sharing}.

\paragraph{Transfer under amplitude damping noise.} To test generalization
beyond depolarizing noise, we evaluate transfer under amplitude damping
($p{=}0.001$) on 6-qubit $\behtwo$, comparing a from-scratch baseline against
transfer at $\epsilon{=}1.00$ and $\epsilon{=}0.55$.
\begin{table}[h!]
\footnotesize
\centering
\caption{\small\textbf{Transfer under amplitude damping noise (6-qubit
$\behtwo$, $p=0.001$).} Buffer transfer improves both energy error and CNOT
count relative to the from-scratch baseline, with the largest gains at reduced
initial exploration ($\epsilon=0.55$).}
\label{tab:amplitude_damping}
\begin{tabular}{lcccc}
\toprule
Setting & $\epsilon$ & Error (Ha) & CNOT & ROT \\
\midrule
No transfer & 1.00 & $8.45\times10^{-5}$ & 29 & 16 \\
\textbf{Transfer}    & 1.00 & $5.81\times10^{-5}$ & \textbf{26} & 15 \\
\textbf{Transfer}    & 0.55 & $\mathbf{5.77\times10^{-5}}$ & $27$ & $\mathbf{10}$ \\
\bottomrule
\end{tabular}
\end{table}
Transfer reduces error from $8.45\times10^{-5}$~Ha to $5.77\times10^{-5}$~Ha
at $\epsilon{=}0.55$ ($31.7\%$ improvement) and CNOT count by $11.3\%$,
confirming the mechanism is effective under non-unital noise and not an
artifact of the depolarizing model used elsewhere.

\paragraph{Score-weight sensitivity.} Equation~\ref{eq:score_transfer} uses a
single fixed weighting throughout. Recomputing $S$ under six alternative weightings at 6- and 12-qubit scale shows in Appendix~\ref{appndx:score_sensitivity} that the transfer benefit becomes markedly less weight-sensitive as system size grows; at 12 qubits, transfer consistently improves speed and CNOT count but not final energy accuracy, a distinction we make explicit rather than implying a uniform accuracy benefit.

\paragraph{Buffer initialization and exploration interact.} In Appendix~\ref{appndx:transfer_ablation} a $2\times2$ ablation study varying buffer initialization (empty vs. transferred, for three noise strengths) and $\epsilon$ on 6- and 8-qubit tasks shows the preferred $\epsilon$ flips depending on transfer source; under the strongest 6-qubit noise
($p_2{=}0.005$), $\epsilon{=}1.0$ gives an order-of-magnitude error improvement over $\epsilon{=}0.55$. Buffer transfer's benefit is thus conditional on its interaction with exploration rate, not uniformly additive.

\paragraph{Why the transfer works.}
Noiseless and noisy tasks share identical state/action spaces, so their
optimal value functions differ only by a bounded perturbation. Classical
transfer analyses~\cite{taylor2009transfer} show high-value source
trajectories remain informative under modest dynamics shifts; Bellman-alignment
theory~\cite{chai2026optimistic} formalizes this via one-step Bellman
mismatch, motivating noiseless-collected transitions as approximately
consistent Bellman targets for the noisy task.
\vspace{-6pt}
\subsection{ReaPER$+$ generalizes beyond quantum domain}
\vspace{-8pt}
{To confirm the annealing mechanism is not quantum-specific,
we benchmark all three replay strategies on
\textbf{LunarLander-v3}~\cite{towers2024gymnasium} with identical DQN agents
and schedule ($\omega_{\min}{=}0.1\to\omega_{\max}{=}0.7$,
$T_{\mathrm{ann}}{=}5{\times}10^4$ steps). Appendix~\ref{app:lunarlander} we show that over 6 seeds, ReaPER$+$ achieves the highest mean success
rate ($0.638\pm0.037$ vs.\ $0.524\pm0.102$ for ReaPER and $0.503\pm0.056$ for
PER) and a $3\%$ AUC gain over both baselines; steps to first solve are
comparable given high seed variance, confirming the PER$\to$ReaPER annealing
principle generalizes across reward regimes.}
\vspace{-6pt}
\section{Conclusion}
\vspace{-8pt}
We introduced a \textit{previously unexplored} replay-buffer engineering framework for quantum circuit optimization, comprising three complementary components. ReaPER$+$ improves sample efficiency and circuit compactness through annealed PER$\rightarrow$ReaPER prioritization, achieving up to $4\times$ greater sample efficiency than the strongest off-policy baselines while matching on-policy solution quality with up to $32\times$ fewer interactions. OptCRLQAS reduces wall-clock time per episode by up to $67.5\%$ through the amortization of expensive quantum-classical evaluations and, when combined with ReaPER$+$ and tensor network warm-starts, extends curriculum-based architecture search to 20 qubits as a structural proof-of-concept. Lightweight buffer transfer reduces the steps to chemical accuracy by up to $88\%$ and the final energy error by up to $90\%$ relative to from-scratch noisy baselines, without network-weight sharing or reward relabeling. These gains strengthen with system size and generalize beyond depolarizing noise to amplitude-damping channels.

Across quantum compiling and QAS benchmarks, our results show that \textit{how experience is stored, sampled, and transferred} is a primary algorithmic lever for sample-efficient, noise-aware quantum circuit optimization. A classical validation on LunarLander-v3 further confirms that ReaPER$+$'s annealing principle is domain-agnostic: with identical agents and schedules, it improves asymptotic success rate by up to $26.8\%$ over PER and $21.8\%$ over fixed ReaPER, while providing a $3\%$ AUC gain over both baselines (see Appendix~\ref{app:lunarlander}). Limitations (including future directions) and broader impact of our work are discussed in Appendix~\ref{appndx:limitations} and Appendix~\ref{appndx:broader_impact} respectively.

\subsection*{Acknowledgment}
SF and AK gratefully acknowledge financial support by QDNL via the National Growth Fund KAT-1 program. We also acknowledge the use of computational resources of the DelftBlue supercomputer, provided by Delft High Performance Computing Centre (\url{https://www.tudelft.nl/dhpc}).

\bibliographystyle{unsrt}
\bibliography{neurips_ref}

\newpage
\appendix
\section{Limitations and future work}
\label{appndx:limitations}

\paragraph{Limitations.}
(1) All quantum experiments use a fixed DQN/DDQN backbone; whether ReaPER$+$'s annealing advantage persists under off-policy actor-critic agents~\cite{haarnoja2018soft} remains an open question; pure on-policy methods fall outside this scope as they do not employ replay buffers. (2) The replay-buffer transfer scheme assumes shared state and action spaces between source and target tasks, which may not hold when porting across substantially different hardware topologies or gate sets. (3) The classical ML benchmarks are restricted to \texttt{LunarLander-v3}, a
relatively low-dimensional discrete-action environment. Evaluating
ReaPER$+$ on high-dimensional benchmarks such as the Atari 2600
suite~\cite{bellemare2013arcade} (e.g., \textit{Pong},
\textit{Breakout}, \textit{Seaquest}) or continuous control domains
such as MuJoCo~\cite{todorov2012mujoco} remains an important open
direction. Finally, (4) all experiments are conducted on classical simulators; execution on real quantum hardware, where device-specific noise, qubit connectivity constraints, and gate calibration drift introduce challenges beyond those captured by depolarizing noise models, remains an important direction for future work, depending on hardware access.

\paragraph{Future work.}
Immediate extensions include: {(1)} combining OptCRLQAS with
tensor network warm-starts~\cite{kundu2025tensorrl} to further reduce search cost at $20$-qubit scale; {(2)} hardware-aware replay, where buffered circuits are ranked using limited QPU evaluations before transfer to realistic noise models; and {(3)} prioritized generative replay~\cite{wang2025prioritized} to synthesize high-value transitions for richer buffer initialization under depolarizing noise.

\section{Broader Impact}
\label{appndx:broader_impact}

This work advances quantum circuit optimization by introducing a previously unexplored replay-buffer engineering framework for deep reinforcement learning (RL). By improving the sample efficiency, computational efficiency, and noise awareness of RL agents for circuit synthesis and quantum architecture search, the proposed methods may lower the practical cost of optimizing circuits for near-term quantum hardware. Beyond quantum circuit optimization, the central idea---that the design of experience storage, sampling, and transfer can substantially affect learning performance---may be relevant to a broader range of sequential decision-making and combinatorial optimization problems.

\paragraph{Positive impacts}
By reducing circuit depth and gate count under realistic noise conditions, our approach can help make meaningful quantum computations more feasible on noisy intermediate-scale quantum devices. This may support progress in near-term applications such as quantum chemistry, materials modeling, combinatorial optimization, and quantum simulation, where reducing resource requirements is essential for obtaining useful results before fault-tolerant hardware becomes widely available.

The replay-buffer engineering principles developed here may also extend beyond quantum computing. In particular, annealed replay prioritization, computational reuse through replay, and lightweight experience transfer could be applicable to RL problems in which environment interactions or reward evaluations are expensive. Potential examples include robotic control with costly real-world data collection, scientific experiment design, molecular and materials discovery, autonomous laboratory systems, scheduling and resource allocation, network control, and energy-system optimization. In such settings, improving the usefulness of previously collected trajectories may reduce the number of expensive interactions, simulations, or experiments needed to learn effective policies.

The framework may further be relevant to transfer and continual-learning settings, where an agent must adapt to related tasks, changed operating conditions, or noisy deployment environments. For example, buffer transfer could potentially support adaptation across related physical systems, device calibrations, manufacturing conditions, or optimization instances without requiring full retraining or direct sharing of model parameters.

\paragraph{Negative impacts}
Training RL agents with large prioritized replay buffers on high-fidelity quantum simulators can demand substantial GPU, memory, and storage resources. This may widen the gap between well-funded and under-resourced research groups, particularly when experiments require extensive hyperparameter sweeps, repeated random seeds, or large-scale simulation. The energy consumption associated with prolonged GPU training and simulator-based evaluations also raises environmental concerns.

Moreover, methods that improve optimization efficiency can be applied broadly, including in domains where automated decision-making may have uneven social, economic, or environmental consequences. For example, applications in scheduling, resource allocation, or autonomous optimization should be evaluated carefully for fairness, safety, robustness, and the consequences of errors under distribution shift. In quantum computing specifically, improved circuit optimization could contribute to both beneficial scientific applications and, in the longer term, capabilities with dual-use implications. These concerns are not unique to the present work, but they motivate transparent reporting of computational cost, reproducible evaluation protocols, and domain-specific safeguards when deploying RL-based optimization systems.

We encourage future work to develop lightweight replay-buffer variants, report energy and hardware costs, investigate methods that remain effective under constrained computational budgets, and adopt green-computing practices and infrastructure. We also encourage evaluation of replay-buffer engineering in real-world high-cost decision-making domains, with appropriate safety, fairness, and human-oversight considerations.

\section{Preliminaries}\label{appndx:preliminaries}

Here we discuss the preliminaries based on which the methods (in Section~\ref{sec:methods}) are based on. This discussion includes a thorough description of the state-of-the-art replay buffer. It also includes the encoding methods through which we represent the action space, reward functions and the RL-state.

\subsection{Replay buffers in deep reinforcement learning}

Experience replay is a central component of off-policy deep RL because it determines which past transitions are revisited during optimization. Uniform replay~\cite{mnih2013playing}, hindsight experience
replay~\cite{andrychowicz2017hindsight}, prioritized experience replay~\cite{schaul2015prioritized}, and more recent reliability-aware variants~\cite{pleiss2026reliabilityadjusted} provide increasingly structured ways to trade off coverage, informativeness, and target reliability. In quantum optimization, however, replay buffers have largely been treated as fixed implementation choices rather than as a primary object of algorithmic design.

To formalize the replay mechanisms studied in this work, we consider an episodic reinforcement-learning setting. An episode is a sequence $D=\{C_t\}_{t=1}^{n}$ of transitions, where each transition $C_t=(S_t,A_t,R_t,S_{t+1},d_t)$ records the state $S_t\in\mathcal{S}$, action $A_t\in\mathcal{A}$, scalar reward $R_t$, next state $S_{t+1}$, and terminal flag $d_t\in\{0,1\}$, which equals $1$ if the episode ends at step $t$ and $0$ otherwise. The agent maintains an online Q-network $Q_\theta$ with parameters $\theta$ and a periodically updated target network $Q_{\bar{\theta}}$ with parameters $\bar{\theta}$, and $\gamma\in(0,1]$ denotes the discount factor. For each stored transition, we define four quantities that replay strategies use to assign sampling priorities. The {TD target} $Y_t$ is the bootstrapped return estimate used to update $Q_\theta$:
\begin{equation}
    Y_t = R_t + \gamma(1-d_t)\max_{a'} Q_{\bar{\theta}}(S_{t+1},a').
\end{equation}
The {TD error} $\delta_t$ measures how far the current value estimate deviates from this target:
\begin{equation}
    \delta_t = Y_t - Q_{\theta}(S_t,A_t),
\end{equation}
and its magnitude $\delta_t^+ = |\delta_t|$ serves as a proxy for how much a transition still has to teach the agent. The {true value error} $e_t$ measures the deviation of the current estimate from the optimal Q-function $Q^\star$:
\begin{equation}
    e_t = Q_\theta(S_t,A_t) - Q^\star(S_t,A_t).
\end{equation}
Finally, the {target bias} $\varepsilon_t$ captures the error introduced by the target network itself:
\begin{equation}
    \varepsilon_t = Q_{\bar{\theta}}(S_t,A_t) - Q^\star(S_t,A_t).
\end{equation}

Replay strategies differ in how they assign sampling preference to stored transitions. For each transition $C_i$ in the replay buffer, let $\Psi_i$ denote its unnormalized replay priority and $\mu_i = \Psi_i / \sum_{j=1}^N \Psi_j$ the corresponding sampling
probability. Different replay rules are then specified by different choices of $\Psi_i$.

\paragraph{Uniform replay (Vanilla).}
A uniform replay buffer samples all stored transitions with equal probability.
If the current buffer contains \(\{C_i\}_{i=1}^{N}\), then
\begin{equation}
    \Psi_i^{\mathrm{uni}} = 1,
    \qquad
    \mu_i^{\mathrm{uni}} = \frac{1}{N}.
    \label{eq:vanilla_replay_eq}
\end{equation}

\paragraph{Hindsight experience replay (HER).}
HER augments the replay buffer by relabeling goals from achieved future 
states within the same trajectory~\cite{andrychowicz2017hindsight}. The 
key idea is that even a failed trajectory, one that did not reach the 
intended goal $g$, can be treated as a successful one with respect to a 
different goal $g'$, typically the state actually reached. Concretely, 
given a goal-conditioned transition $(S_t,A_t,R_t,S_{t+1},d_t,g)$, HER 
adds an additional relabeled copy to the buffer,
\begin{equation}
    \big(S_t,A_t,R_t'(g'),S_{t+1},d_t'(g'),g'\big),
\end{equation}
where $g'$ is the relabeled goal and $R_t'(g')$, $d_t'(g')$ are the 
reward and terminal flag recomputed with respect to $g'$. Unless stated 
otherwise, HER samples uniformly from the augmented buffer. HER is 
utilized in the quantum compilation task in Ref.~\cite{moro2021quantum}.

\paragraph{Prioritized experience replay (PER).}
PER samples transitions according to their absolute TD error
\cite{schaul2015prioritized}:
\begin{equation}
    \Psi_i^{\mathrm{PER}} = (\delta_i^+)^\alpha,
    \qquad
    \mu_i^{\mathrm{PER}}
    = \frac{\Psi_i^{\mathrm{PER}}}{\sum_{j=1}^{N}\Psi_j^{\mathrm{PER}}},
    \label{eq:PER_sample}
\end{equation}
where $\alpha>0$ controls the strength of prioritization.

\paragraph{Reliability-adjusted prioritized experience replay (ReaPER).}
ReaPER was recently introduced in ref.~\cite{pleiss2026reliabilityadjusted}. It discounts transitions whose targets are unreliable due to large downstream TD errors. For an episode $D=\{C_t\}_{t=1}^{n}$, the reliability score is
\begin{equation}
    \mathcal{R}_t
    =
    1-\frac{\sum_{i=t+1}^{n}\delta_i^+}{\sum_{i=1}^{n}\delta_i^+},
    \label{eq:realibility_score}
\end{equation}
and the corresponding priorities are
\begin{equation}
    \Psi_t^{\mathrm{ReaPER}} = \mathcal{R}_t^{\omega}(\delta_t^+)^{\alpha},
    \qquad
    \mu_t^{\mathrm{ReaPER}}
    =
    \frac{\Psi_t^{\mathrm{ReaPER}}}{\sum_{i=1}^{n}\Psi_i^{\mathrm{ReaPER}}},
    \label{eq:ReaPER_sample}
\end{equation}
where $\omega \in [0,1]$ controls the strength of reliability weighting ($\omega=0$ recovers PER and $\omega=1$ corresponds to fully reliability-adjusted replay). Thus, relative to PER, high-error and low-reliability transitions are down-weighted.

\subsection{Reinforcement learning for quantum optimization}

Reinforcement learning has been applied to both quantum compilation and parameterized quantum circuit design. Existing work includes RL-based gate-sequence synthesis~\cite{ruiz2025quantum, fosel2021quantum, kremer2024practical}, architecture search for molecular ground state preparation~\cite{ostaszewski2021reinforcement}, and hardware-aware variational circuit construction~\cite{patel2024curriculum}, demonstrating that RL can discover nontrivial circuit structures across a range of quantum tasks. At the same time, prior methods typically operate under substantial sample and compute demands, especially when each environment step requires an expensive variational optimization or when the target setting includes hardware noise. To tackle this, we formulate both quantum compilation and quantum architecture search as
Markov decision processes in which an agent sequentially constructs a quantum circuit. At time step $t$, the agent observes $S_t\in\mathcal{S}$, selects $A_t\in\mathcal{A}$, receives reward $R_t$, and transitions to $S_{t+1}$. Episodes terminate when a target threshold is reached or when a maximum circuit length is exceeded.

\subsubsection{RL-state and action for quantum compilation}\label{sec:state_action_compiling}

In quantum compilation, the agent aims to approximate a target unitary
$U_{\mathrm{tar}}$ by appending gates from a finite base $\mathcal{G}$.
Following Ref.~\cite{moro2021quantum}, the system is initialized in the 
all-zero state $\ket{0}^{\otimes n}$, and the circuit unitary after $t$ 
steps is
    $U_t = \prod_{j=1}^{t} A_j$,
so that the prepared state at step $t$ is $U_t\ket{0}^{\otimes n}$. The 
observation is based on the residual unitary
\begin{equation}
    O_t = U_t^\dagger U_{\mathrm{tar}},\label{eq:compiling_rl_state}
\end{equation}
whose real and imaginary entries are provided to the agent.

\paragraph{1-qubit compiling}
For 1-qubit compiling, the chosen gateset base is
\begin{equation}
    \mathcal{G}^{(1)}_{\mathrm{rot}}
    =
    \left\{
    R_i\!\left(\theta_j\right) \;\middle|\; i \in \{x,y,z\},\; \theta_j \in \left\{+\frac{\pi}{128}, -\frac{\pi}{128}\right\}
    \right\},\label{eq:compiling_small_rotations}
\end{equation}
where $R_i(\theta_j)$ denotes a single-qubit rotation by angle $\theta_j$ about axis $i$. The rotation angle is thus restricted to a binary set of small fixed increments $\pm\pi/128$, making each action a minimal discrete rotation rather than a continuous parameter choice. 

In an alternate setting, we also consider the Harrow-Recht-Chuang (HRC)~\cite{harrow2002efficient} efficient universal basis,
    $V_1 =
    \frac{1}{\sqrt{5}}
    \begin{pmatrix}
    1 & 2i\\
    2i & 1
    \end{pmatrix}$,
    $V_2 =
    \frac{1}{\sqrt{5}}
    \begin{pmatrix}
    1 & 2\\
    -2 & 1
    \end{pmatrix}$,
    $V_3 =
    \frac{1}{\sqrt{5}}
    \begin{pmatrix}
    1+2i & 0\\
    0 & 1-2i
    \end{pmatrix}$,
with action space
\begin{equation}
    \mathcal{A}^{(1)}_{\mathrm{HRC}} = \{V_1,V_2,V_3\}.\label{eq:compiling_hrc_gateset}
\end{equation}

\paragraph{2-qubit compiling.}
For the 2-qubit setting, we use the action space:
\begin{equation}
    \mathcal{A}^{(2)}=\{R_z(\theta_j) \otimes \mathbb{I}, \mathbb{I} \otimes R_z(\theta_j), XX = XX\!\left(\pm\frac{\pi}{128}\right), YY = YY\!\left(\pm\frac{\pi}{128}\right)\},\label{eq:compiling_2q_gateset}
\end{equation}

\subsubsection{RL-state and action for quantum architecture search}\label{sec:state_action_qas}

For quantum architecture search (QAS), the agent searches over parameterized quantum circuits to minimize a task-dependent cost, typically the expectation value~\cite{cerezo2021variational}. Following CRLQAS~\cite{patel2024curriculum}, the state is represented by a
tensor-based binary encoding of the circuit:
\begin{equation}
    \mathcal{S}_{\mathrm{QAS}}
    =
    \left\{
    (\mathbf{E}_t,c_t)
    \;\middle|\;
    \mathbf{E}_t\in\{0,1\}^{N_{\mathrm{depth}}\times N_{\mathrm{wires}}\times N_{\mathrm{gate\_type}}},
    \; c_t\in\mathbb{R}
    \right\},
\end{equation}
where $\mathbf{E}_t$ encodes the circuit structure and $c_t$ summarizes its current performance by calculating the current cost function i.e. the expectation value of Hamiltonian. Using the gate set $\mathcal{G}_{\mathrm{QAS}}=\{\texttt{RX}(\theta_x), \texttt{RY}(\theta_y),\texttt{RZ}(\theta_z),\texttt{CNOT}\}$, the action 
space consists of placing a gate at a particular circuit location. A 
compact description is
\begin{equation}
    \mathcal{A}_{\mathrm{QAS}}
    =
    \left\{
    a=(\ell,i,g)
    \;\middle|\;
    \ell\in\{1,\dots,N_{\mathrm{layers}}\},\;
    i\in\{1,\dots,N_{\mathrm{wires}}\}^{k_g},\;
    g\in\mathcal{G}_{\mathrm{QAS}}
    \right\},\label{eq:qas_gateset}
\end{equation}
where $\ell$ is the layer index, $i$ specifies the qubit(s) the gate acts on, and $k_g$ is the number of qubits required by gate $g$ which isequal to $1$ for single-qubit gates $\{\texttt{RX}, \texttt{RY}, \texttt{RZ}\}$ and $2$ for the two-qubit gate $\texttt{CNOT}$. Illegal actions are masked out during training.

\subsubsection{Reward structure}

In both tasks, an episode ends either when a target threshold is met or when
the maximum number of steps is reached.

\paragraph{Quantum compilation.}
Following Ref.~\cite{moro2021quantum}, we use two reward designs depending on
the gate base. For bases small-rotation (i.e $\pm \pi/128$), we use the dense reward
\begin{equation}
    r_t^{\mathrm{comp}}
    =
    \begin{cases}
        (L-t)+1, & \text{if } d(U_t,U_{\mathrm{tar}})<\varepsilon, \\[4pt]
        -d(U_t,U_{\mathrm{tar}})/L, & \text{otherwise},
    \end{cases}
    \label{eq:reward_small_rotation}
\end{equation}
where $L$ is the maximum episode length and $\varepsilon$ is the target
accuracy threshold. For discrete bases such as HRC, we use the sparse reward
\begin{equation}
    r_t^{\mathrm{comp}}
    =
    \begin{cases}
        0, & \text{if } d(U_t,U_{\mathrm{tar}})<\varepsilon, \\[4pt]
        -1/L, & \text{otherwise}.
    \end{cases}
    \label{eq:reward_hrc}
\end{equation}

\paragraph{Quantum architecture search.}
Here we use the reward used as in \cite{ostaszewski2021reinforcement, patel2024curriculum}.
Let $C_t$ be the optimized variational cost at step $t$, $\xi$ the target
threshold, $C_{\min}$ the desired minimum cost, and $T_s^e$ the maximum number
of steps in episode $e$.
Then
\begin{equation}
    R_t^{\mathrm{QAS}}
    =
    \begin{cases}
        5, & \text{if } C_t < \xi, \\[4pt]
        -5, & \text{if } t \ge T_s^e \text{ and } C_t \ge \xi, \\[6pt]
        \max\!\left(
        \dfrac{C_{t-1}-C_t}{C_{t-1}-C_{\min}},\,-1
        \right), & \text{otherwise}.
    \end{cases}
    \label{eq:reward_qas}
\end{equation}
This reward provides a positive terminal signal for success, a negative
terminal signal for failure, and otherwise a normalized improvement-based
intermediate reward.

\section{Sample complexity and justification for ReaPER+}\label{appndx:reaper+}

Here we provide an argument on why the proposed ReaPER$+$ replay rule can be expected to outperform a fixed replay strategy over the full course of training. Recall that ReaPER$+$ defines the replay priority of transition $t$ at training step $\tau$ as
\begin{equation}
    \Psi_t^{(+,\tau)} = \mathcal{R}_t^{\omega_\tau}(\delta_t^+)^\alpha,
    \qquad
    \mu_t^{(+,\tau)}
    =
    \frac{\Psi_t^{(+,\tau)}}{\sum_i \Psi_i^{(+,\tau)}},
\end{equation}
where $\delta_t^+=|\delta_t|$ is the absolute temporal-difference error,
$\mathcal{R}_t \in [0,1]$ is the reliability score defined in Equation~\ref{eq:realibility_score}, $\alpha>0$ is the prioritization
exponent, and $\omega_\tau \in [0,1]$ is a non-decreasing annealing parameter.
When $\omega_\tau=0$, the scheme reduces to PER, whereas $\omega_\tau=1$
recovers ReaPER.

\begin{proposition}
    Assume that: (i) larger reliability $\mathcal{R}_t$ corresponds to smaller bias in the TD target of transition $t$; (ii) early in training, reliability estimates are noisy and only weakly correlated with true target quality; and (iii) later in training, reliability estimates become more informative. Then a schedule $\omega_\tau$ that starts near $0$ and increases toward $1$ induces a replay distribution that is better aligned with the needs of the learning process than either fixed PER~\cite{schaul2015prioritized} or ReaPER~\cite{pleiss2026reliabilityadjusted} used throughout training.
\end{proposition}

\paragraph{Justification.}
Consider two transitions $i$ and $j$. Their relative sampling probability under
ReaPER$+$ is
\begin{equation}
    \frac{\mu_i^{(\omega)}}{\mu_j^{(\omega)}}
    =
    \frac{\mathcal{R}_i^{\omega}(\delta_i^+)^\alpha}
         {\mathcal{R}_j^{\omega}(\delta_j^+)^\alpha}
    =
    \left(\frac{\mathcal{R}_i}{\mathcal{R}_j}\right)^\omega
    \left(\frac{\delta_i^+}{\delta_j^+}\right)^\alpha.
\end{equation}
If $\mathcal{R}_i > \mathcal{R}_j$, then for any $\omega_2 > \omega_1$ we have
\begin{equation}
    \frac{\mu_i^{(\omega_2)}}{\mu_j^{(\omega_2)}}
    >
    \frac{\mu_i^{(\omega_1)}}{\mu_j^{(\omega_1)}}.
\end{equation}
Hence increasing $\omega$ monotonically shifts replay mass toward transitions with higher reliability while preserving the PER-style dependence on TD error.
\begin{figure}[h!]
    \centering
    \caption{\small \textbf{ReaPER+ progressively concentrates buffer mass toward higher-fidelity transitions} (fidelity $\geq 0.95$) while retaining broader early-training coverage, consistent with its annealed transition from PER-like exploration to ReaPER-like reliability-aware sampling. PER maintains broader low-fidelity coverage throughout training, while ReaPER shows intermediate concentration behavior.}
    \includegraphics[width=0.8\linewidth]{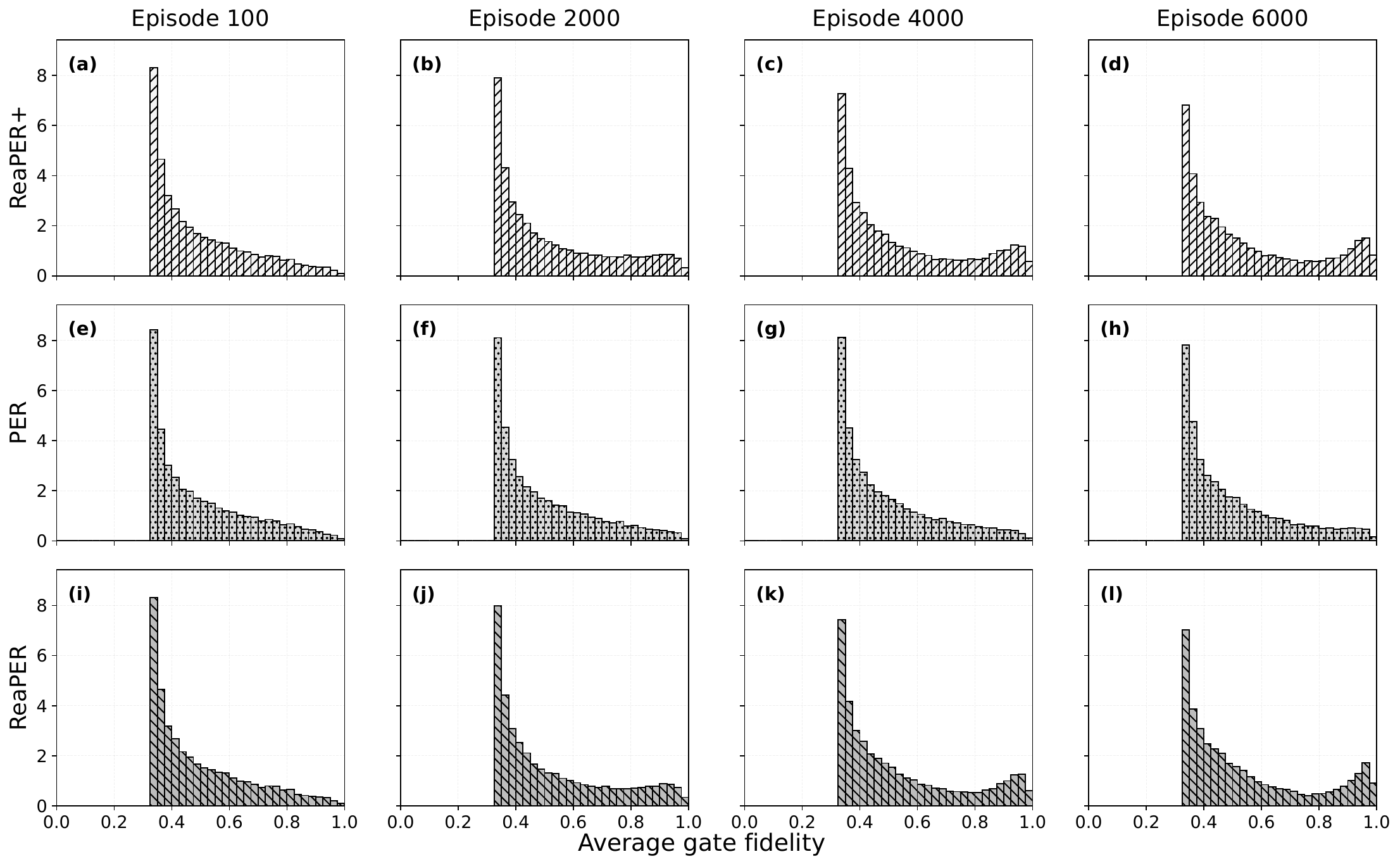}
    \label{fig:buffer_compiling_1q_poc}
\end{figure}

This monotonicity explains the role of the annealing schedule. At the beginning of training, the value function is inaccurate and the resulting reliability estimates can be unstable. In this regime, setting $\omega_\tau \approx 0$ prevents the learner from overcommitting to a noisy reliability signal and recovers the PER~\cite{schaul2015prioritized} behavior that emphasizes large TD error transitions. Later in training, once the critic becomes more accurate, the reliability score carries more information about whether a large TD error corresponds to a genuinely useful correction or to a misleading target. Increasing $\omega_\tau$ then down-weights high-error but low-reliability transitions and shifts the replay distribution toward more trustworthy updates.

Equivalently, one may examine the log-priority
\begin{equation}
    \log \Psi_t^{(+,\tau)}
    =
    \omega_\tau \log \mathcal{R}_t + \alpha \log \delta_t^+.
\end{equation}
The annealing coefficient $\omega_\tau$ controls the contribution of $\log R_t$ relative to $\log \delta_t^+$. Early in training, the replay rule is dominated by the TD error term; later, the reliability term gradually regularizes prioritization by suppressing transitions whose downstream targets are less trustworthy. Therefore, ReaPER+ inherits the strong early-learning behavior of PER and the late-stage stability of ReaPER.

The argument above does not claim universal dominance for every environment or every annealing schedule. Rather, it shows that under mild assumptions on the quality of reliability estimates across training, ReaPER+ provides a natural continuation method between two useful replay regimes. As shown in Figure~\ref{fig:buffer_compiling_1q_poc}, during the 1-qubit compiling task, ReaPER$+$ progressively shifts the fidelity distribution toward higher-fidelity, more reliable replay samples significantly faster than both PER and ReaPER.

\subsection{The sample complexity bound for ReaPER$+$}
\label{appndx:regret-bound}

We establish a formal sample complexity guarantee for ReaPER$+$ under
tabular Q-learning. The main result shows that annealing the replay
distribution from PER-like to ReaPER-like sampling provably reduces the
asymptotic noise floor of the Q-learning iterates compared to fixed PER.

\begin{theorem}[Conditional Variance Reduction]
\label{thm:sample-complexity}
Consider tabular Q-learning on a fixed dataset $\mathcal{D} =
\{C_i\}_{i=1}^N$ under the ReaPER$^+$ annealed sampling distribution
with linear schedule $\omega_t = \min\!\bigl(\omega_{\min} +
\tfrac{\omega_{\max}-\omega_{\min}}{T_{\mathrm{ann}}}t,\,
\omega_{\max}\bigr)$. Under Assumptions~(1)-(4), Assumption~\ref{assump:rva},
and the MLR condition of Lemma~\ref{lem:variance}, the expected supremum
error after $T$ updates satisfies

\begin{align}
  \mathbb{E}[\|Q_T - Q^\star\|_\infty]
  &\leq
  \underbrace{(1-\alpha(1-\gamma))^T \|Q_0 - Q^\star\|_\infty}_{\textrm{initialization decay}}
  \notag\\
  &\quad+
  \underbrace{\frac{\alpha}{1-\gamma}
    \Bigl(
      \frac{T_{\mathrm{ann}}}{T}\,\sigma_{\mathrm{PER}}
      +
      \frac{T-T_{\mathrm{ann}}}{T}\,\sigma_{\mathrm{ReaPER}(\omega_{\max})}
    \Bigr)}_{\textrm{noise floor}}
  \notag\\
  &\quad+
  \underbrace{O(\zeta(T_{\mathrm{ann}})) + O(B)}_{\textrm{irreducible error}}.
  \label{eq:main-bound}
\end{align}

where $B \leq \frac{2\gamma R_{\max}}{1-\gamma}$ is the bootstrap bias
constant and $\zeta(T_{\mathrm{ann}})$ captures the reliability score
estimation error during the annealing transition. When reliability scores $\mathcal{R}_i$ successfully down-weight high-variance transitions, $\sigma_{\mathrm{ReaPER}(\omega_{\max})} \leq \sigma_{\mathrm{PER}}$, so 2nd term in Equation~\ref{eq:main-bound} for ReaPER$+$ is strictly smaller than that of fixed PER, yielding a lower asymptotic noise floor.
\end{theorem}

\paragraph{Scope of the result.} This guarantee is conditional on
Assumption~\ref{assump:rva} and the MLR property holding for the empirical joint distribution of $(\mathcal{R}_i, \delta_i)$; it should be read as an idealized intuition motivating the annealed schedule rather than an unconditional guarantee for arbitrary replay buffers. Verifying the MLR condition directly requires estimating a conditional likelihood ratio over the full joint distribution of $(\mathcal{R}_i, \delta_i)$, which is computationally prohibitive for buffers containing millions of transitions, and we do not attempt this verification here. Even the weaker, directly testable implication of Assumption~\ref{assump:rva}, that $\mathbb{E}[\delta_i^2 \mid \mathcal{R}_i \geq r]$ is decreasing in $r$,
would require dedicated logging and binning infrastructure not currently part of our training pipeline. We therefore present
Theorem~\ref{thm:sample-complexity} as a conditional theoretical result whose premises are plausible given the design of the reliability score in Equation~\ref{eq:realibility_score}, but whose empirical validation on logged buffers remains an open direction for future work.

The three terms in Equation~\eqref{eq:main-bound} have the following
interpretation. The 1st term decays geometrically to zero and is independent of the replay strategy. The 2nd term is the dominant term at convergence and is directly controlled by the sampling distribution that is where ReaPER$+$ improves over PER. Finally, the 3rd term collects irreducible errors from bootstrap bias and reliability estimation, both of which are bounded constants independent of $T$.

The remainder of this section provides a complete proof of
Theorem~\ref{thm:sample-complexity}. We first state the setting and assumptions, then establish the bias bound in Lemma~\ref{lem:bias-bound}, then the one-step contraction in Lemma~\ref{lem:contraction}, followed by the variance bound in Lemma~\ref{lem:variance}, and finally the unrolled recursion with annealing
in Lemma~\ref{lem:unrolling}.

The agent maintains an online Q-network $Q_\theta$ and a target network
$Q_{\bar{\theta}}$, following the notation of
Section~\ref{appndx:preliminaries}. The TD target and TD error for
transition $C_i = (S_i, A_i, R_i, S_{i+1}, d_i)$ are
\begin{equation}
  Y_i = R_i + \gamma(1-d_i)\max_{a'} Q_{\bar{\theta}}(S_{i+1}, a'),
  \qquad
  \delta_i = Y_i - Q_\theta(S_i, A_i),
\end{equation}
The reliability score $\mathcal{R}_i \in [r_{\min}, 1]$ is defined as in
Equation~(\ref{eq:realibility_score}), and the ReaPER$+$ sampling
probability follows the annealed priority
\begin{equation}
  \Psi_i^{(t)} = \mathcal{R}_i^{\omega_t}(\delta_i^+)^\alpha,
  \qquad
  \mu_i^{(t)} = \frac{\Psi_i^{(t)}}{\sum_{j=1}^N \Psi_j^{(t)}},
\end{equation}
recovering PER at $\omega_t = 0$ and full ReaPER at $\omega_t =
\omega_{\max}$. The importance-sampling corrected update for the $Q$-value is
\begin{equation}
  Q_{t+1}(S_i, A_i)
  = Q_t(S_i, A_i) + \alpha\, w_i^{(t)}\bigl(Y_i - Q_t(S_i, A_i)\bigr),
\end{equation}
where $w_i^{(t)} = \tfrac{1}{N \mu_i^{(t)}}$ is the importance-sampling weight.

\paragraph{Analysis-implementation weight mismatch.} The analysis above uses
fully bias-corrected importance-sampling weights $w_i^{(t)} = 1/(N\mu_i^{(t)})$.
Our deep RL implementation instead follows standard PER
practice~\cite{schaul2015prioritized}, using $\beta$-annealed weights
$w_i^{(t)} = (N\mu_i^{(t)})^{-\beta_t}$ with $\beta_t$ annealed from
$\beta_0 < 1$ toward $1$ over training, since applying full correction from
initialization is known to destabilize early optimization when $\mu_i^{(t)}$
estimates are still unreliable. Theorem~\ref{thm:sample-complexity} therefore
provides qualitative support for the direction of the annealing schedule
rather than an exact guarantee for the specific weighting scheme used in the
deployed algorithm.

Throughout, we denote by $R_{\max} = \max_{s,a}|r(s,a)|/(1-\gamma)$ the
supremum of the discounted return, by $\Delta_{\max} =
\sup_{i,t}|\delta_i^{(t)}|$ the uniform upper bound on TD errors, and by
$r_{\min} > 0$ the minimum reliability score in the buffer. To realize the Theorem~\ref{thm:sample-complexity} we require the
following conditions. (1)~\textit{Boundedness:} the Q-function satisfies $|Q_t(s,a)| \leq R_{\max}$ at all times, which is the natural bound
implied by discounted cumulative rewards~\cite{gosavi2006boundedness};
(2)~\textit{Bounded TD error:} the temporal-difference errors satisfy
$|\delta_i| \leq \Delta_{\max}$, ensuring the variance analysis in
Lemma~\ref{lem:variance} is finite; (3)~\textit{Step size:} the learning
rate satisfies $\alpha \in \bigl(0, \frac{1}{1-\gamma}\bigr)$, which
guarantees the contraction factor $(1-\alpha(1-\gamma)) \in (0,1)$ in
Lemma~\ref{lem:contraction}; in practice we further restrict to $\alpha <
\frac{1}{2\gamma}$ to keep the bias constant $B$
finite~\cite{fan2020theoretical}; (4)~\textit{Reliability bounded:} the reliability scores satisfy
$\mathcal{R}_i \in [r_{\min}, 1]$, preventing any transition from receiving
zero sampling probability and ensuring the importance-sampling weights
$w_i^{(t)}$ remain well-defined and finite.

\begin{assumption}[Reliability-Variance Alignment]
\label{assump:rva}
For all thresholds $r \in [r_{\min}, 1]$,
\begin{equation}
  \mathbb{E}[\delta_i^2 \mid \mathcal{R}_i \geq r]
  \;\leq\;
  \mathbb{E}[\delta_i^2 \mid \mathcal{R}_i < r].
\end{equation}
That is, transitions with higher reliability scores have, on average, no
larger squared TD error than transitions with lower reliability scores.
This is the minimal condition under which reliability-weighted sampling can
be expected to reduce variance rather than merely redistribute it.
\end{assumption}

\begin{lemma}[Bias bound, in expectation]
\label{lem:bias-bound}
Under Assumption~(1), the bias introduced by the fixed target network
$Q_{\bar\theta}$ satisfies
\begin{equation}
  B \;\coloneqq\; \mathbb{E}_{i\sim P}\Bigl[\bigl|w_i^{(t)}\bigl(Y_i - Q_t(S_i,A_i)\bigr)
  - \bigl((TQ_t)(S_i,A_i) - Q_t(S_i,A_i)\bigr)\bigr|\Bigr]
  \;\leq\; \frac{2\gamma R_{\max}}{1-\gamma}.
\end{equation}
\end{lemma}

\begin{proof}
By definition of $Y_i$ and the Bellman operator $T$, the reward terms cancel and
\[
Y_i - (TQ_t)(S_i,A_i) = \gamma\Bigl(\max_{a'} Q_{\bar\theta}(S_{i+1},a') - \mathbb{E}_{S_{i+1}}\bigl[\max_{a'} Q_t(S_{i+1},a')\bigr]\Bigr),
\]
so $|Y_i - (TQ_t)(S_i,A_i)| \le \gamma\|Q_{\bar\theta}-Q_t\|_\infty \le \gamma\Bigl(\|Q_{\bar\theta}-Q^\star\|_\infty + \|Q_t-Q^\star\|_\infty\Bigr) \le \frac{2\gamma R_{\max}}{1-\gamma}$
pointwise for every $i$, by the same argument as before. Taking expectation
over $i\sim P$ preserves this bound by Jensen's inequality, giving
$B \le \mathbb{E}_{i\sim P}\bigl[\gamma\|Q_{\bar\theta}-Q_t\|_\infty\bigr] \le \frac{2\gamma R_{\max}}{1-\gamma}$.
This expectation-based statement is also the form actually used in
Lemma~\ref{lem:contraction} below, which only ever invokes $B$ inside an
expectation over $Q_{t+1}$, so this correction additionally removes an
internal inconsistency between Lemmas D.2 and D.3.
\end{proof}

\begin{lemma}[One-step contraction]
\label{lem:contraction}
Under Assumptions~(1)--(3) and Lemma~\ref{lem:bias-bound},
\begin{equation}
  \|\mathbb{E}[Q_{t+1} \mid Q_t] - Q^\star\|_\infty
  \leq (1-\alpha(1-\gamma))\|Q_t - Q^\star\|_\infty + \alpha B.
\end{equation}
\end{lemma}

\begin{proof}
By the importance-sampling update rule and linearity of expectation,
\begin{equation}
  \mathbb{E}[Q_{t+1}\mid Q_t]
  = Q_t + \alpha\,\mathbb{E}\bigl[w_i^{(t)}(Y_i - Q_t)\mid Q_t\bigr].
\end{equation}
Writing $w_i^{(t)}(Y_i - Q_t) = (TQ_t - Q_t) + \bigl[w_i^{(t)}(Y_i -
Q_t) - (TQ_t - Q_t)\bigr]$ and taking expectation,
\begin{equation}
  \mathbb{E}[Q_{t+1}\mid Q_t]
  = Q_t + \alpha(TQ_t - Q_t)
    + \alpha\,\mathbb{E}\bigl[w_i^{(t)}(Y_i - Q_t)
      - (TQ_t - Q_t)\mid Q_t\bigr],
\end{equation}
where the last term is bounded in absolute value by $B$ by
Lemma~\ref{lem:bias-bound}. Subtracting $Q^\star$ and using the identity
$Q_t + \alpha(TQ_t - Q_t) - Q^\star = (1-\alpha)(Q_t - Q^\star) +
\alpha(TQ_t - Q^\star)$,
\begin{equation}
  \mathbb{E}[Q_{t+1}\mid Q_t] - Q^\star
  = (1-\alpha)(Q_t - Q^\star) + \alpha(TQ_t - Q^\star) + \alpha\,\eta_t,
\end{equation}
where $\|\eta_t\|_\infty \leq B$. Taking $\|\cdot\|_\infty$, applying
the triangle inequality, and using the $\gamma$-contraction
$\|TQ_t - Q^\star\|_\infty \leq \gamma\|Q_t -
Q^\star\|_\infty$~\cite{tsitsiklis1994asynchronous} together with
$TQ^\star = Q^\star$,
\begin{equation}
  \|\mathbb{E}[Q_{t+1}\mid Q_t] - Q^\star\|_\infty
  \leq \bigl((1-\alpha) + \alpha\gamma\bigr)\|Q_t - Q^\star\|_\infty
     + \alpha B
  = (1-\alpha(1-\gamma))\|Q_t - Q^\star\|_\infty + \alpha B,
\end{equation}
where $(1-\alpha(1-\gamma)) \in (0,1)$ by Assumption~(3).
\end{proof}

\begin{lemma}[Variance bound, conditional]
\label{lem:variance}
Under Assumptions~(2) and~(4), the variance of the importance-sampling
weighted TD update satisfies
\begin{equation}
  \sigma_t^2 \;\coloneqq\; \mathbb{E}\bigl[(w_i^{(t)}\delta_i)^2\bigr]
  = \frac{1}{N^2}\sum_{i=1}^N \frac{\delta_i^2}{\mu_i^{(t)}}.
\end{equation}
Suppose further that Assumption~\ref{assump:rva} holds and that the sampling
distribution satisfies the Monotone Likelihood Ratio (MLR) property with
respect to the reliability scores:
\begin{equation}
  \frac{P(\mathcal{R}_i = r \mid \delta_i^2 \text{ is small})}
       {P(\mathcal{R}_i = r \mid \delta_i^2 \text{ is large})}
  \quad\text{is strictly increasing in } r.
\end{equation}
Then $\sigma_t^2$ is non-increasing in $\omega_t$, with maximum at
$\omega_t = 0$ (PER) and minimum at $\omega_t = \omega_{\max}$ (full ReaPER).
Without the MLR condition, increasing $\omega_t$ need not reduce $\sigma_t^2$,
since $\sum_i 1/\mu_i^{(t)}$ is minimized by the \emph{uniform} distribution;
reliability-weighting only reduces variance relative to PER when it
concentrates mass specifically on the low-error transitions identified by
Assumption~\ref{assump:rva}.
\end{lemma}

\begin{proof}
Since index $i$ is drawn according to $\mu^{(t)}$, the law of total
expectation and the substitution $w_i^{(t)} = \frac{1}{N\mu_i^{(t)}}$ give
\begin{equation}
  \mathbb{E}\bigl[(w_i^{(t)}\delta_i)^2\bigr]
  = \sum_{i=1}^N \mu_i^{(t)}
    \left(\frac{\delta_i}{N\mu_i^{(t)}}\right)^{\!2}
  = \frac{1}{N^2}\sum_{i=1}^N \frac{\delta_i^2}{\mu_i^{(t)}}.
\end{equation}
The weights are well-defined by Assumption~(4). For monotonicity, note that
under $\Psi_i^{(t)} = \mathcal{R}_i^{\omega_t}(\delta_i^+)^\alpha$, increasing
$\omega_t$ reweights $\mu^{(t)}$ toward transitions with higher
$\mathcal{R}_i$. The MLR condition guarantees this reweighting is
\emph{targeted}: it implies that, for any threshold, the conditional
odds of high $\mathcal{R}_i$ given small $\delta_i^2$ dominate the
conditional odds given large $\delta_i^2$, so raising $\omega_t$
provably shifts probability mass from high-$\delta_i^2$ transitions to
low-$\delta_i^2$ transitions rather than redistributing mass arbitrarily.
Combined with Assumption~\ref{assump:rva}, this mass shift strictly
decreases $\sum_i \delta_i^2/\mu_i^{(t)}$ in expectation, since probability
is moved away from terms with large $\delta_i^2/\mu_i^{(t)}$ toward terms
with small $\delta_i^2$. Hence $\sigma_t^2$ is non-increasing in
$\omega_t$, with the extremes at $\omega_t=0$ and $\omega_t=\omega_{\max}$
following by substitution.
\end{proof}

\begin{lemma}[Unrolled recursion with annealing]
\label{lem:unrolling}
Under Lemmas~\ref{lem:contraction} and~\ref{lem:variance}, iterating
the one-step recursion over $T$ steps and applying the annealing
partition yields

\begin{align}
  \mathbb{E}[\|Q_T - Q^\star\|_\infty]
  &\leq
  (1-\alpha(1-\gamma))^T \|Q_0 - Q^\star\|_\infty
  \notag\\
  &\quad+
  \frac{\alpha}{1-\gamma}
    \Bigl(
      \frac{T_{\mathrm{ann}}}{T}\,\sigma_{\mathrm{PER}}
      + \frac{T-T_{\mathrm{ann}}}{T}\,\sigma_{\mathrm{ReaPER}(\omega_{\max})}
    \Bigr)
  \notag\\
  &\quad+
  O(\zeta(T_{\mathrm{ann}})) + O(B).
\end{align}

\end{lemma}

\begin{proof}
Combining Lemmas~\ref{lem:contraction} and~\ref{lem:variance} via
standard stochastic approximation arguments~\cite{tsitsiklis1994asynchronous},
\begin{equation}
  \mathbb{E}[\|Q_{t+1} - Q^\star\|_\infty]
  \leq (1-\alpha(1-\gamma))\,\mathbb{E}[\|Q_t - Q^\star\|_\infty]
  + \alpha\sqrt{\sigma_t^2} + \alpha B.
\end{equation}
Let $\rho \coloneqq 1-\alpha(1-\gamma) \in (0,1)$. Iterating this
recursion over $T$ steps,
\begin{equation}
  \mathbb{E}[\|Q_T - Q^\star\|_\infty]
  \leq \rho^T\|Q_0 - Q^\star\|_\infty
    + \alpha\sum_{t=0}^{T-1}\rho^{T-1-t}\bigl(\sqrt{\sigma_t^2} + B\bigr).
\end{equation}
Bounding the geometric sum $\sum_{t=0}^{T-1}\rho^{T-1-t} \leq
\frac{1}{1-\rho} = \frac{1}{\alpha(1-\gamma)}$ and replacing the
geometrically-weighted average by a uniform Cesàro average at the cost
of a constant factor,
\begin{equation}
  \mathbb{E}[\|Q_T - Q^\star\|_\infty]
  \leq \rho^T\|Q_0 - Q^\star\|_\infty
    + \frac{\alpha}{1-\gamma}\cdot\frac{1}{T}\sum_{t=1}^T\sqrt{\sigma_t^2}
    + O(B).
\end{equation}
The linear schedule partitions training into a PER-like phase ($t \leq
T_{\mathrm{ann}}$, variance $\sigma_{\mathrm{PER}}$) and a ReaPER-like
phase ($t > T_{\mathrm{ann}}$, variance $\sigma_{\mathrm{ReaPER}(\omega_{\max})}$)
by Lemma~\ref{lem:variance}. The time-averaged noise therefore satisfies
\begin{equation}
  \frac{1}{T}\sum_{t=1}^T\sqrt{\sigma_t^2}
  \leq \frac{T_{\mathrm{ann}}}{T}\,\sigma_{\mathrm{PER}}
     + \frac{T-T_{\mathrm{ann}}}{T}\,\sigma_{\mathrm{ReaPER}(\omega_{\max})}
     + O(\zeta(T_{\mathrm{ann}})),
\end{equation}
where $\zeta(T_{\mathrm{ann}})$ accounts for the estimation error of
$\mathcal{R}_i$ during the transition region. Substituting into the
unrolled bound yields the statement.
\end{proof}

Theorem~\ref{thm:sample-complexity} follows immediately by combining
Lemmas~\ref{lem:bias-bound}--\ref{lem:unrolling}. The annealing
schedule, controlled by $T_{\mathrm{ann}} = T/2$ in all experiments
(Section~\ref{sec:methods}), ensures PER-like exploration robustness
early in training and variance reduction once the reliability estimates
$\mathcal{R}_i$ have stabilized, directly explaining the empirical gains
reported in Tables~\ref{tab:2q_haar},~\ref{tab:6q_comparison} and
Figure~\ref{fig:haar1q-summary}.

\subsection{ReaPER$+$ implementation details}
\label{appndx:reaper_impl}

Algorithm~\ref{alg:reaper_plus_impl} documents the exact replay-buffer and training-step logic used in our implementation, as a companion to the idealized formulation in Section~\ref{sec:methods} and Appendix~\ref{appndx:reaper+}.
\begin{algorithm}[h!]
\footnotesize
\caption{ReaPER$+$ replay buffer and training step implementation}
\label{alg:reaper_plus_impl}
\KwIn{SumTree buffer $\mathcal{T}$ (capacity $N$); exponent $\alpha$; fixed/annealed $\omega$; IS-exponent $\beta$ (start $\beta_0$, horizon $F_\beta$); $\epsilon=10^{-6}$; networks $Q_\theta, Q_{\bar\theta}$; sync interval $K$}

\Proc{\FPush{$s_t,a_t,r_t,s_{t+1},d_t$, episode $e$}}{
  $\hat\delta_t \leftarrow \max(|r_t|,0.1)$\tcp*[r]{reward-based warm-start, not a true TD error}
  Insert at max tree priority; append index to episode $e$'s list\;
  $\Sigma_e \leftarrow \Sigma_e + |\hat\delta_t|$\;
  \If{$d_t=1$}{
    \FCompute{$e$}\;
  }
}

\Proc{\FCompute{$e$}}{
  \eIf{$\Sigma_e \le \epsilon$}{
    $p_i \leftarrow (|\hat\delta_i|+\epsilon)^\alpha$ for all $i \in e$\;
  }{
    \ForEach{position $k$ in $e$}{
      $d_k \leftarrow \sum_{j>k}|\hat\delta_j|$\;
      $\mathcal{R}_k \leftarrow \FClip(1-d_k/\Sigma_e,\,0,\,1)$\;
      $p_k \leftarrow \mathcal{R}_k^{\omega}(|\hat\delta_k|+\epsilon)^\alpha$\;
    }
  }
}

\Proc{\FSample{$B$}}{
  $\beta \leftarrow \min(1,\,\beta_0+\mathrm{frame}\cdot(1-\beta_0)/F_\beta)$\;
  Stratify $[0,\mathcal{T}.\mathrm{total}]$ into $B$ segments; draw and retrieve one transition per segment\;
  $\mu_i \leftarrow p_i/\mathcal{T}.\mathrm{total}$\;
  $w_i \leftarrow (N\mu_i)^{-\beta}$, normalized by $\max_j w_j$\;
  \Return{batch, indices $I=\{idx_i\}$, weights $W=\{w_i\}$}\;
}

\Proc{\FUpdate{$I, \{\delta_i\}$}}{
  \ForEach{$(idx_i,\delta_i)$}{
    $e \leftarrow$ episode of $idx_i$\;
    $\Sigma_e \leftarrow \Sigma_e + |\delta_i| - |\hat\delta_i^{\mathrm{old}}|$\;
    Overwrite stored TD error at $idx_i$ with $\delta_i$\;
    \eIf{$\Sigma_e \le \epsilon$}{
      $\mathcal{R}_i \leftarrow 1$\;
    }{
      $d_i \leftarrow \max(0,\,\Sigma_e-|\delta_i|)$\tcp*[r]{approximation, not exact suffix-sum recompute}
      $\mathcal{R}_i \leftarrow \FClip(1-d_i/\Sigma_e,\,0,\,1)$\;
    }
    $p_i \leftarrow \mathcal{R}_i^{\omega}(|\delta_i|+\epsilon)^\alpha$\;
  }
}

\Proc{\FTrain{$B$}}{
  \If{$|\mathcal{T}|<B$}{
    \Return{None}\;
  }
  \If{step $\bmod\ K=0$}{
    $\bar\theta \leftarrow \theta$\;
  }
  $(s,a,r,s',d,I,W) \leftarrow$ \FSample{$B$}\;
  $a^\star \leftarrow \arg\max_{a'} Q_\theta(s')$\;
  $Y \leftarrow r+\gamma_{\mathrm{layer}}(1-d)Q_{\bar\theta}(s')[a^\star]$\;
  $\delta \leftarrow |Y-Q_\theta(s)[a]|$\;
  \FUpdate{$I,\delta$}\;
  $\mathcal{L} \leftarrow \mathrm{SmoothL1}(W\odot Q_\theta(s)[a],\, W\odot Y)$\tcp*[r]{weights scale both sides, see note}
  $\theta \leftarrow \theta-\eta\nabla_\theta\mathcal{L}$\;
  Anneal $\epsilon_{\mathrm{greedy}}$\;
}
\end{algorithm}
Three implementation choices deviate from the idealized formulation and are noted inline in the algorithm: (i)~new transitions are seeded with a reward-based priority proxy $\max(|r_t|,0.1)$ rather than a true TD error, with exact reliability computed only once the episode completes; (ii)~priority updates between episode boundaries approximate the downstream TD sum as $\Sigma_e - |\delta_i|$ rather than recomputing the exact suffix sum of Equation~\ref{eq:PER_sample}; and (iii)~the importance-sampling weight is applied to both the prediction and the target before computing the SmoothL1 loss, which effectively strengthens the weighting relative to the standard convention of weighting the loss term alone.

\section{Replay buffer transfer under source-target task 
similarity}\label{appndx:buffer-transfer}

We justify why replay-buffer transfer from a noiseless source environment to a 
corresponding noisy target environment can accelerate learning in quantum 
optimization. Let $\mathcal{M}_{\mathrm{src}}$ and $\mathcal{M}_{\mathrm{tgt}}$ 
denote two Markov decision processes with the same state and action spaces, 
where the target task differs from the source task by noise in the transition 
dynamics, reward evaluations, or both. Let $\mathcal{B}_{\mathrm{src}}$ be a 
replay buffer collected in the source task, and let the target replay buffer be 
initialized as
\begin{equation}
    \mathcal{B}_{\mathrm{tgt}}^{(0)}
    \leftarrow
    \mathcal{T}(\mathcal{B}_{\mathrm{src}}),
\end{equation}
where $\mathcal{T}$ is a transfer operator that copies compatible transitions 
from source to target. The following proposition provides a structured 
justification for this scheme under source-target task similarity. It is not 
intended as a formal convergence theorem under general function approximation 
— rigorous sample-complexity bounds in that setting remain an open problem — 
but rather as a precise statement of the conditions under which transfer is 
expected to help, with a constructive argument for each mechanism.

\begin{proposition}
Assume the source and target tasks share the same state and action spaces 
$\mathcal{S}$ and $\mathcal{A}$ (\textit{shared structure}); that the target 
task is a bounded perturbation of the source task in the sense that there exist 
constants $\varepsilon_r, \varepsilon_p \geq 0$ such that for all 
$(s, a) \in \mathcal{S} \times \mathcal{A}$,
\begin{align}
    |r^{\mathrm{tgt}}(s,a) - r^{\mathrm{src}}(s,a)| &\leq \varepsilon_r,\label{eq:transfer_eq1_appndx} \\
    \left\|\mathbb{E}_{P^{\mathrm{tgt}}}\!\left[\max_{a'} Q(s',a')\right] 
    - \mathbb{E}_{P^{\mathrm{src}}}\!\left[\max_{a'} Q(s',a')\right]
    \right\| &\leq \varepsilon_p,
    \label{eq:transfer_eq2_appndx}
\end{align}
(\textit{bounded task shift}); that the Q-function is uniformly bounded, 
$\|Q\|_\infty \leq Q_{\max} < \infty$, which holds under bounded rewards 
$|r| \leq R_{\max}$ and $\gamma \in (0,1)$ since 
$Q_{\max} \leq R_{\max}/(1-\gamma)$ (\textit{bounded Q-function}); and that 
high-value source trajectories remain informative, though not necessarily 
optimal, in the target task, with source transitions concentrated in 
high-reward regions of $\mathcal{S} \times \mathcal{A}$ overlapping with 
high-value regions of the target task (\textit{informative source 
trajectories}). Then the Bellman targets of corresponding source and target 
transitions differ by a bounded amount,
\begin{equation}
    |Y^{\mathrm{tgt}}(s,a) - Y^{\mathrm{src}}(s,a)| \leq 
    \varepsilon_r + \gamma\varepsilon_p,
\end{equation}
and initializing the target replay buffer with transferred source transitions 
improves early target learning relative to training from an empty buffer via 
three mechanisms: \emph{(a)} reduced Bellman target mismatch, \emph{(b)} 
improved initial buffer coverage over promising regions of circuit space, and 
\emph{(c)} self-correction as on-target experience progressively replaces 
stale source transitions.
\end{proposition}

\paragraph{Justification.}
Let the source and target Bellman targets for a transition $(s,a)$ be defined 
as in Equation~\ref{eq:transfer_eq1_appndx}-Equation~\ref{eq:transfer_eq2_appndx}. Under the \textit{bounded task 
shift} assumption, the triangle inequality yields
\begin{equation}
    |Y^{\mathrm{tgt}}(s,a) - Y^{\mathrm{src}}(s,a)|
    \leq \varepsilon_r + \gamma\varepsilon_p,
\end{equation}
as suggested by classical simulation-lemma arguments and recent 
Bellman-alignment analyses for transfer RL~\cite{chai2026optimistic}. Thus a 
transition informative for the source task remains approximately informative 
for the target task whenever the source-target shift is moderate. In our 
quantum optimization setting, both assumptions are naturally satisfied: the 
\textit{informative source trajectories} assumption holds because noiseless and 
noisy environments share identical circuit-building structure, so gate sequences 
that reduce energy error under noiseless simulation identify the same 
architecturally productive regions of circuit space under depolarizing noise where noise acts as a bounded stochastic perturbation on the cost landscape rather than altering which gate combinations are structurally meaningful. Finally, the 
\textit{bounded task shift} assumption is satisfied because depolarizing noise with 1-qubit strength $p_1$ and 2-qubit strength $p_2$ introduces a perturbation that is bounded yet grows with system size, which is precisely why the transfer advantage grows with qubit count (Figure~\ref{fig:6q_buffer_transfer_matrix}-Figure~\ref{fig:12q_buffer_transfer_matrix}): as the 
noiseless-to-noisy gap widens, the warm-start from source experience becomes 
proportionally more valuable relative to from-scratch exploration.

Beyond bounded target mismatch, transferred source transitions provide two further benefits. First, \textit{improved coverage}: at the start of noisy training, an empty buffer contains no information about which gate sequences or architectural motifs are promising, whereas a transferred buffer is already concentrated in regions of state-action space that were useful in the source task. Since noiseless and noisy tasks share the same circuit-building structure, these transitions provide a more informative initial replay distribution than random target experience alone, allowing the learner to update from semantically meaningful trajectories before independently rediscovering them under noise. Second, \textit{self-correction}: as the agent interacts with the target 
environment, new target-domain transitions are added to the buffer and progressively shift the replay distribution toward the true noisy task. Transfer therefore does not lock the agent into the source domain; it supplies a warm start that is gradually refined by on-target experience. In prioritized replay schemes this correction is especially effective, since target transitions with 
larger TD errors or higher reliability automatically gain greater replay probability, accelerating adaptation.

\section{Classical RL validation: LunarLander-v3}
\label{app:lunarlander}

The three contributions presented in the main text ReaPER$+$, OptCRLQAS,
and replay-buffer transfer, are motivated by and evaluated on quantum circuit optimization tasks. To verify that the annealing schedule underlying Reaper$+$ is not specific to the sparse, long-horizon reward structure of quantum environments, we provide a supplementary validation on \texttt{LunarLander-v3}~\cite{towers2024gymnasium}, a well-established continuous-control benchmark from classical deep RL withdense rewards and a standard solved threshold of $+200$ cumulative return.
We stress that this experiment is \emph{not} intended as a primary result; its sole purpose is to confirm that the PER$\rightarrow$ReaPER annealing idea generalizes beyond the quantum domain.
\begin{figure}[h!]
  \centering
  \caption{\small
    \textbf{LunarLander-v3 validation of ReaPER$+$, averaged over 6 seeds.}
    (\textit{Left}) Rolling success rate (1000-episode window) for ReaPER$+$,
    PER, and fixed ReaPER. (\textit{Middle}) Normalized cumulative-return AUC;
    ReaPER$+$ retains a $+3\%$ AUC advantage over both baselines. (\textit{Right})
    Steps to first solve (mean $\pm$ std): PER $439\mathrm{k}\pm59\mathrm{k}$,
    ReaPER$+$ $431\mathrm{k}\pm133\mathrm{k}$, ReaPER $421\mathrm{k}\pm30\mathrm{k}$
    -- these differences are within one standard deviation given ReaPER$+$'s
    high seed-to-seed variance. All methods use identical DQN agents, only the
    replay mechanism differs. Fills/error bars denote 1-sigma.}
  \includegraphics[width=\linewidth]{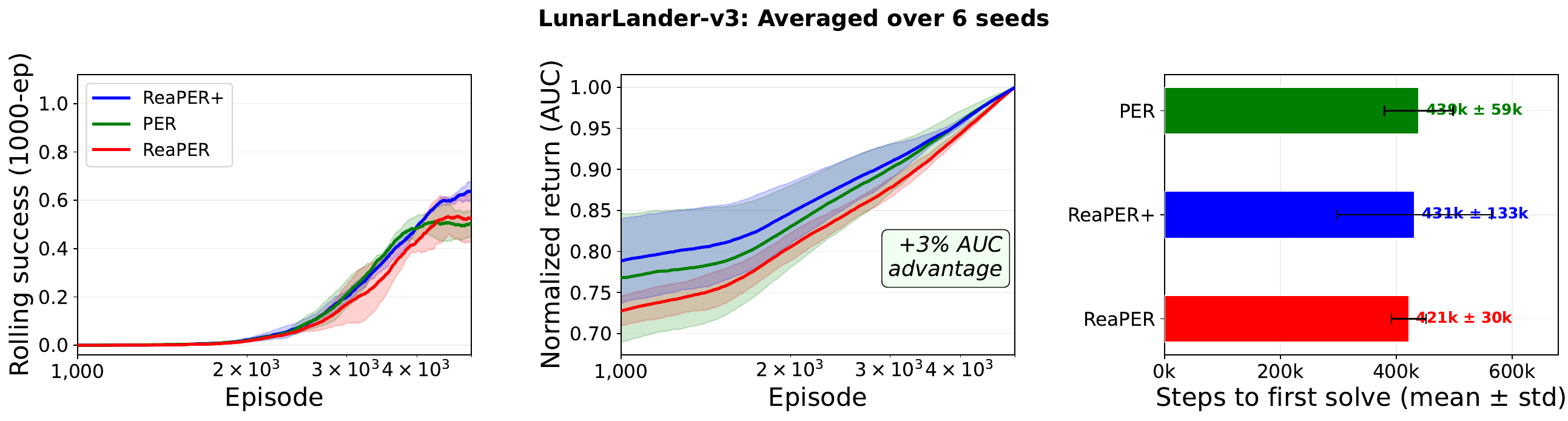}
  \label{fig:lunarlander}
\end{figure}

We benchmark three replay strategies using identical DQN agents:
\begin{itemize}
\item \textbf{PER}~\cite{schaul2015prioritized}: prioritization by absolute TD error, $\alpha{=}0.6$, $\beta_0{=}0.4$ annealed to $1.0$ over $5{\times}10^4$ frames.
\item \textbf{ReaPER}~\cite{pleiss2026reliabilityadjusted}: reliability-adjusted prioritization with fixed $\omega{=}0.4$, $\alpha{=}0.4$.
\item \textbf{ReaPER$+$ (ours)}: annealed $\omega$ schedule
        $\omega_{\min}{=}0.1\!\to\!\omega_{\max}{=}0.7$ over $T_{\mathrm{ann}}{=}5{\times}10^4$
        steps (identical schedule to the quantum compiling experiments),
        $\alpha{=}0.4$.
\end{itemize}
All agents are trained for $5{,}000$ episodes on \texttt{LunarLander-v3}
with $\gamma{=}0.99$, $\varepsilon$-greedy exploration ($\varepsilon_0{=}1.0\!\to\!\varepsilon_{\min}{=}0.05$, decay $0.9995$ per episode),
learning rate $10^{-3}$, batch size $64$, replay capacity $10^5$,
and target-network synchronization every $10$ episodes. Four gradient updates are performed per episode after a $20$-episode warm-up.
The same agent architecture and training code used for the quantum experiments are employed without modification; only the environment is swapped. Results are averaged over $6$ independent random seeds (extended from $3$ seeds in the original submission), reported in Table~\ref{tab:lunarlander_seeds}.

Figure~\ref{fig:lunarlander} summarizes three complementary metrics.
\emph{Left}: the $1000$-episode rolling success rate (fraction of episodes
exceeding the $+200$ solved threshold). ReaPER$+$ sustains the highest
asymptotic success rate ($0.638\pm0.037$) compared with fixed ReaPER
($0.524\pm0.102$) and PER ($0.503\pm0.056$). \emph{Middle}: the normalized
area under the reward curve (AUC), computed as the per-episode running mean
of the cumulative return, shifted and normalized to $[0,1]$. ReaPER$+$
retains a $+3\%$ AUC advantage over both baselines under the expanded seed
count, down from the $+9\%$ reported under $3$ seeds -- indicating that part
of the original margin was attributable to seed variance, though a smaller,
more reliable advantage persists. \emph{Right}: steps to first solve. Here
PER and fixed ReaPER reach first solve in nominally fewer average steps
($439\mathrm{k}$ and $421\mathrm{k}$, respectively) than ReaPER$+$
($431\mathrm{k}$), reversing the direction reported under $3$ seeds; however,
ReaPER$+$'s standard deviation on this metric ($\pm133\mathrm{k}$) is
$3$--$4\times$ larger than either baseline's, so this ordering is not
statistically meaningful. We report it transparently rather than omitting it,
and treat ReaPER$+$'s consistently higher success rate as the more robust
signal of the annealing schedule's benefit in this domain.

\begin{table}[h]
\small
\centering
\caption{\small\textbf{LunarLander-v3 validation over 6 seeds.} Mean $\pm$
standard deviation of success rate and steps to first solve.}
\label{tab:lunarlander_seeds}
\begin{tabular}{lccc}
\toprule
Metric & PER & ReaPER & ReaPER$+$ \\
\midrule
Success rate & $0.503\pm0.056$ & $0.524\pm0.102$ & $\mathbf{0.638\pm0.037}$ \\
Steps        & $438541.7\pm59143.9$ & $\mathbf{421304.0\pm29782.3}$ & $431427.3\pm132823.6$ \\
\bottomrule
\end{tabular}
\end{table}

These results are consistent with the mechanism described in
Section~\ref{sec:methods} and Appendix~\ref{appndx:reaper+}:
early in training, $\omega \approx 0$ recovers PER-style exploration;
as value estimates stabilize, rising $\omega$ down-weights unreliable
high-error transitions and concentrates replay on trustworthy updates.
This two-phase behavior is beneficial regardless of whether rewards are
sparse and episodic (quantum compilation) or dense and continuous
(LunarLander), supporting the view that ReaPER$+$ is a general-purpose
annealing strategy rather than a domain-specific heuristic; the expanded
seed count refines the magnitude of this advantage without changing its
direction.

\begin{table}[h!]
  \centering
  \caption{Hyperparameters for the LunarLander-v3 validation experiment.
           Shared parameters apply to all three replay strategies; method-specific
           parameters are listed in the lower block.}
  \label{tab:lunarlander_hparam}
  \small
  \begin{tabular}{lc}
    \toprule
    \textbf{Parameter} & \textbf{Value} \\
    \midrule
    \multicolumn{2}{l}{\textit{Shared (all methods)}} \\
    Network architecture    & 2 $\times$ 128 hidden units, SELU \\
    Discount $\gamma$       & $0.99$ \\
    Learning rate           & $10^{-3}$ \\
    Batch size              & $64$ \\
    Replay capacity         & $10^5$ \\
    $\varepsilon$ schedule  & $1.0 \to 0.05$, decay $0.9995$/episode \\
    Gradient updates/ep.    & $4$ (after $20$-episode warm-up) \\
    Training episodes       & $5000$ \\
    Solved threshold        & $+200$ cumulative return \\
    \midrule
    \multicolumn{2}{l}{\textit{PER}} \\
    $\alpha$                & $0.6$ \\
    $\beta_0$               & $0.4$, annealed to $1.0$ over $5{\times}10^4$ frames \\
    \midrule
    \multicolumn{2}{l}{\textit{ReaPER}} \\
    $\alpha$                & $0.4$ \\
    $\omega$                & $0.4$ (fixed) \\
    $\beta_0$               & $0.4$, annealed to $1.0$ over $5{\times}10^4$ frames \\
    \midrule
    \multicolumn{2}{l}{\textit{ReaPER$+$}} \\
    $\alpha$                & $0.4$ \\
    $\omega_{\min}$         & $0.1$ \\
    $\omega_{\max}$         & $0.7$ \\
    $T_{\mathrm{ann}}$      & $5{\times}10^4$ steps \\
    $\beta_0$               & $0.4$, annealed to $1.0$ over $5{\times}10^4$ frames \\
    \bottomrule
  \end{tabular}
\end{table}

\section{1-qubit sample significance analysis}
\label{appndx:1q_significance}

The circuit-length comparisons in Figure~\ref{fig:haar1q-summary} are aggregated over a small number of independent training seeds ($n=5$), which limits the precision with
which we can estimate the true population mean gate count for each method. To quantify this uncertainty rigorously, rather than relying on point-estimate comparisons alone, we compute the standard error of the mean and construct approximate 95\% confidence intervals using the $t$-distribution with $df=n-1=4$.
Table~\ref{tab:1q_significance} reports the resulting mean gate count, standard deviation, and confidence interval for PER, fixed ReaPER, and ReaPER$+$ at each tolerance level. As expected given the small sample size, the intervals are wide
and overlap substantially across methods, so we do not claim statistical
\begin{table}[h!]
\small
\centering
\caption{\small\textbf{Sample significance analysis of circuit length.} Mean gate
count $\pm$ standard deviation and approximate 95\% confidence interval ($df=4$)
for the small-rotation basis compiling task.}
\label{tab:1q_significance}
\begin{tabular}{llcc}
\toprule
Tol & Method & Gate count & 95\% CI (approx.) \\
\midrule
\multirow{3}{*}{0.99}
  & PER            & $79.00 \pm 48.32$  & $[19.0, 139.0]$ \\
  & ReaPER         & $90.48 \pm 61.78$  & $[13.8, 167.2]$ \\
  & \textbf{ReaPER+} & $\mathbf{75.72 \pm 48.36}$  & $\mathbf{[15.7, 135.8]}$ \\
\midrule
\multirow{3}{*}{0.999}
  & PER            & $85.93 \pm 53.22$  & $[19.9, 152.0]$ \\
  & ReaPER         & $89.21 \pm 61.98$  & $[12.3, 166.2]$ \\
  & \textbf{ReaPER+} & $\mathbf{70.88 \pm 50.74}$  & $\mathbf{[7.9, 133.9]}$ \\
\midrule
\multirow{3}{*}{0.9999}
  & PER            & $128.51 \pm 53.13$ & $[62.6, 194.5]$ \\
  & ReaPER         & $123.94 \pm 62.83$ & $[45.9, 201.9]$ \\
  & \textbf{ReaPER+} & $\mathbf{74.09 \pm 41.13}$  & $\mathbf{[23.0, 125.2]}$ \\
\bottomrule
\end{tabular}
\end{table}
significance in the classical hypothesis-testing sense. However, the analysis is
informative in a different way: it shows that the sample mean for ReaPER$+$ is
consistently lower than PER and fixed ReaPER at all three tolerances, and that this
directional shift persists even under the most conservative (widest) interval
estimate, which supports the qualitative trend reported in the main text and
motivates evaluating ReaPER$+$ on a larger seed budget in future work.

\section{1-qubit compiling with HRC gateset}\label{appndx:1q_HRC_compiling}
For a direct comparison with Ref.~\cite{moro2021quantum}, we evaluate the discrete HRC basis~\cite{harrow2002efficient} using the action space and sparse reward defined in Equation~\ref{eq:compiling_hrc_gateset} and Equation~\ref{eq:reward_hrc}. Table~\ref{tab:hrc_1q_haar} shows that ReaPER$+$ reaches $100\%$ success with a mean fidelity of $0.995$ and the shortest average circuit length ($14.30\pm7.89$ gates), achieving this at $1.56\times10^6$ steps. This is slightly earlier than fixed ReaPER
($1.60\times10^6$ steps), clearly earlier than PER ($2.10\times10^6$ steps), and substantially earlier than our tuned HER baseline ($5.50\times10^6$ steps),
while also producing shorter circuits than all three. 
\begin{table}[h!]
\centering
\caption{\small\textbf{1-qubit compiling of Haar-random targets using the HRC basis.} Success rate, mean fidelity, and circuit length for different replay strategies. All methods except the original HER baseline~\cite{moro2021quantum} reach 100\% success. ReaPER$+$ yields the shortest circuits ($14.30 \pm 7.89$ gates) and reaches peak performance
${\sim}24\%$, ${\sim}26\%$, and ${\sim}72\%$ faster than fixed ReaPER, PER, and tuned HER, respectively.}
\label{tab:hrc_1q_haar}
\vspace{2mm} 
\small
\begin{tabular}{lcccc}
\toprule
\textbf{Method} & \textbf{Success (\%)\,$\uparrow$} & \textbf{Fidelity\,$\uparrow$} & \textbf{Mean length\,$\downarrow$} & \textbf{At step\,$\downarrow$} \\
\midrule
\textbf{ReaPER$+$ (Ours)} & $\mathbf{100.0}$ & $\mathbf{0.995}$ & $\mathbf{14.30 \pm 7.89}$  & $\mathbf{1.56\times10^6}$ \\
ReaPER~\cite{pleiss2026reliabilityadjusted}  & $100.0$          & $0.995$          & $20.34 \pm 16.16$           & $1.60\times10^6$  \\
PER~\cite{schaul2015prioritized}             & $100.0$          & $0.995$          & $19.54 \pm 14.61$           & $2.10\times10^6$  \\
HER (tuned)                                  & $100.0$          & $0.992$          & $28.60 \pm 21.58$           & $5.50\times10^6$  \\
HER (Moro et al.~\cite{moro2021quantum})     & $95.0$           & $0.990$          & $35$                        & --              \\
\bottomrule
\end{tabular}
\end{table}
In contrast, the original unoptimized HER implementation from Moro et al.~\cite{moro2021quantum} plateaus at $95\%$ accuracy. Note that throughout the remainder of this
paper, any reference to the HER baseline indicates our tuned implementation rather than the original formulation of Ref.~\cite{moro2021quantum}, ensuring all comparisons are made against the strongest possible version of the baseline.
\section{Target dataset for 2-qubit compiling task}
For the compiling experiments with the 1-qubit small-rotation basis and the 2-qubit gate set, we generate target unitaries synthetically by sampling random circuits from the corresponding elementary gate library. Concretely, a target unitary is constructed by first sampling a circuit length uniformly at random and then composing gates drawn uniformly from the same basis used by the agent during training. This yields a diverse
dataset of reachable targets with varying circuit complexity while ensuring that the target distribution is consistent with the underlying compilation task. For the 2-qubit setting, Algorithm~\ref{alg:2qubit_target}
details this procedure for the gate set
\(\{XX(\pm \pi/128),\, YY(\pm \pi/128),\, R_z(\pm \pi/128)\otimes\mathbb{I},\, \mathbb{I}\otimes R_z(\pm \pi/128)\}\);
the 1-qubit small-rotation targets are generated analogously from their
corresponding 1-qubit rotation basis.
\begin{algorithm}[h!]
\caption{2-qubit target unitary generation.}
\label{alg:2qubit_target}
\KwIn{Gate set $\mathcal{B} = \bigl\{XX(\pm\tfrac{\pi}{128}),\;
      YY(\pm\tfrac{\pi}{128}),\;
      R_z(\pm\tfrac{\pi}{128})\otimes\mathbb{I},\;
      \mathbb{I}\otimes R_z(\pm\tfrac{\pi}{128})\bigr\}$}
\KwOut{Target unitary $U_{\mathrm{tar}} \in \mathrm{SU}(4)$}

Sample circuit length $N \sim \mathrm{Uniform}\bigl(\{6, 7, \ldots, 10^4 - 1\}\bigr)$\;

Initialize $U_{\mathrm{tar}} \leftarrow \mathbb{I}_4$\;

\For{$k = 1$ \KwTo $N$}{
    Sample gate $G_k \sim \mathrm{Uniform}(\mathcal{B})$\;
    $U_{\mathrm{tar}} \leftarrow G_k \cdot U_{\mathrm{tar}}$\;
}
\Return $U_{\mathrm{tar}}$
\end{algorithm}

\section{Benchmarking OptCRLQAS against CRLQAS~\cite{patel2024curriculum}}\label{appndx:optcrlqas_vs_crlqas}

Before we provide the comprehensive benchmark of OptCRLQAS against CRLQAS we discuss how the reward function and and RL-state calculated and populated in the intermediate steps in Section~\ref{appndx:reward_state_optcrlqas} and then the benchmark is provided in Section~\ref{appndx:opt_crlqas_crlqas_benchmark_sub}.

\subsection{Reward and RL-state during intermediate steps}\label{appndx:reward_state_optcrlqas}

In standard CRLQAS, a full quantum-classical evaluation is performed at every 
environment step, so the variational cost $C_t$, the RL-state cost component 
$c_t$, and the reward $R_t^{\mathrm{QAS}}$ (Equation~\ref{eq:reward_qas}) are all updated 
synchronously. OptCRLQAS breaks this synchrony by restricting full evaluations to every $m$ steps. We describe here how rewards and states are populated during the $m-1$ intermediate steps between evaluations.

\paragraph{Cost function evaluation} Let $\tau_{\mathrm{prev}}$ denote the most recent 
step at which $u_\tau = 1$, i.e., the last step at which a full 
quantum-classical evaluation was performed. During all intermediate steps 
$\tau \in (\tau_{\mathrm{prev}},\, \tau_{\mathrm{prev}} + m)$, the variational 
cost is held constant at its cached value:
\begin{equation}
    C_\tau \leftarrow C_{\tau_{\mathrm{prev}}}, \quad u_\tau = 0.
\end{equation}
This cached value is used for both reward computation and state construction 
until the next evaluation is triggered.

\paragraph{State representation} The QAS state $S_\tau = (E_\tau, c_\tau)$ 
(Equation~18) carries the current circuit structure $E_\tau$, which is updated 
at every step as the agent appends gates, and the scalar cost summary 
$c_\tau = C_{\tau_{\mathrm{prev}}}$, which remains frozen between evaluations. 
The agent therefore observes the evolving circuit structure but receives a 
delayed cost signal, analogous to a control system operating with periodic 
rather than continuous feedback.

\paragraph{Reward computation} The intermediate reward follows 
Equation~22 with the cached cost substituted for the true cost:
\begin{equation}
R_\tau^{\mathrm{QAS}} = 
\begin{cases}
5, & \text{if } C_{\tau_{\mathrm{prev}}} < \xi, \\
-5, & \text{if } \tau \geq T_s^e \text{ and } C_{\tau_{\mathrm{prev}}} \geq \xi, \\
\max\!\left(\dfrac{C_{\tau_{\mathrm{prev}}-m} - C_{\tau_{\mathrm{prev}}}}{C_{\tau_{\mathrm{prev}}-m} - C_{\min}},\, -1\right), & \text{otherwise},
\end{cases}
\end{equation}
where $C_{\tau_{\mathrm{prev}}-m}$ is the cost from the evaluation immediately 
preceding the current block. During intermediate steps the improvement-based 
term evaluates to zero, since $C_{\tau_{\mathrm{prev}}-m}$ and 
$C_{\tau_{\mathrm{prev}}}$ are both fixed within the block. The agent therefore 
receives a zero intermediate reward during accumulation steps, with the 
non-trivial learning signal arriving only at evaluation steps. This is 
intentional: rather than rewarding each individual gate placement which 
produces weakly separated TD targets when variational parameters $\theta$ can 
locally compensate single-gate changes — OptCRLQAS credits the outcome of an 
entire block of $m$ edits jointly. As discussed in Section~4.2, this increases 
reward contrast and accelerates value propagation, steering the agent toward 
architecturally meaningful gate combinations rather than single-step local 
optima.

\paragraph{Terminal conditions} Terminal detection ($d_\tau = 1$) is checked 
at every step regardless of the update schedule. If the episode terminates at 
an intermediate step (i.e., $u_\tau = 0$ but $\tau \geq T_s^e$), a final 
evaluation is immediately triggered ($u_\tau$ is overridden to 1) so that the 
terminal reward reflects the true cost of the final circuit rather than a 
stale cached value. This ensures that the terminal transition carries an 
accurate TD target and prevents the agent from exploiting the caching 
mechanism to avoid negative terminal penalties.

\subsection{Benchmark}\label{appndx:opt_crlqas_crlqas_benchmark_sub}

\begin{figure}[h!]
    \centering
    \caption{\small \textbf{Comparative runtime analysis of the baseline CRLQAS versus OptCRLQAS across $6$-, $8$-, and $10$-qubit tasks}. The left panel shows the quantum energy evaluation time (in seconds), and the right panel displays the classical optimization time (in seconds). Results for OptCRLQAS are reported for different replay buffer sizes ($m \in \{3, 5, 7, 10\}$). Error bars denote the standard deviation measured over $3$ to $5$ independent episodes. The utilization of OptCRLQAS yields an average runtime reduction of up to $89\%$ for quantum evaluations and $85\%$ for classical optimizations at $m=10$. Note that the error bars are 1-sigma.}
    \includegraphics[width=\linewidth]{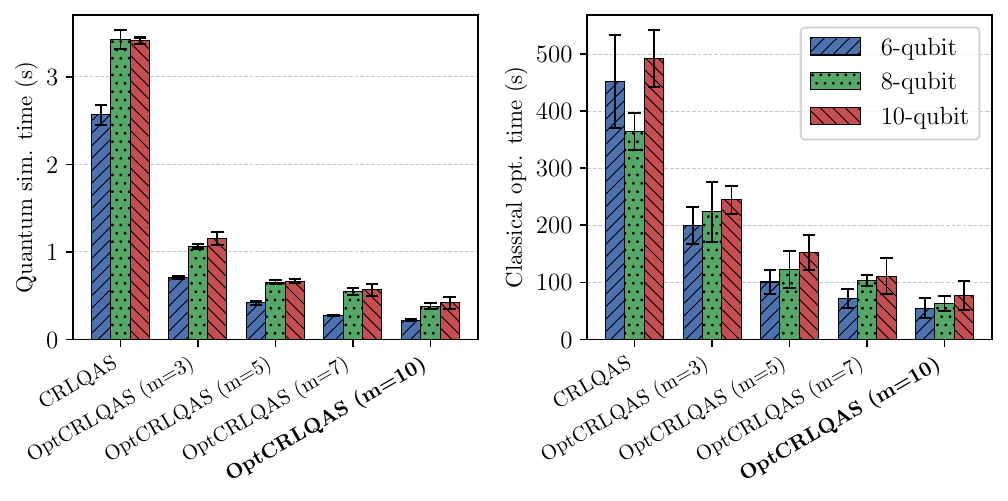}
    \label{fig:optcrlqas_crlqas_compare_6_to_10q}
\end{figure}

To empirically validate the computational efficiency gained by amortizing the architecture update overhead, we benchmarked OptCRLQAS against the baseline CRLQAS framework across $6$-, $8$-, and $10$-qubit configurations. We tracked both the expectation value evaluation and the classical optimization time over multiple episodes. Figure \ref{fig:optcrlqas_crlqas_compare_6_to_10q} illustrates the substantial reduction in computational cost when utilizing OptCRLQAS across varying step intervals ($m \in \{3, 5, 7, 10\}$).

Because CRLQAS triggers a full quantum-classical evaluation at every environment step, its runtime scales poorly as circuit complexity increases. By accumulating local architecture edits and restricting full evaluations to every $m$ steps, OptCRLQAS closely matches the expected runtime reduction factor of $\approx 1/m$. Our empirical results demonstrate an aggressive decline in runtime as $m$ increases. For quantum energy evaluations, OptCRLQAS achieves an average time reduction of $69.2\%$ at $m=3$, scaling up to an impressive $89.3\%$ reduction at $m=10$ across all qubit scales. Similarly, the classical optimization time is heavily mitigated, yielding an average decrease of $48.2\%$ at $m=3$ and up to $84.9\%$ at $m=10$. This benchmarking clearly establishes that accumulating structural modifications before parameter optimization successfully breaks the runtime bottleneck, enabling RL-based quantum architecture search to scale to larger qubit regimes without prohibitive GPU compute requirements.

\begin{table}[h!]
\centering
\small
\caption{\textbf{Comparison between vanilla CRLQAS and OptCRLQAS (\(m=10\)) on 8-qubit H$_2$O over three random seeds}. For OptCRLQAS, the total gate count is computed as \(\texttt{ROT}+{\texttt{CNOT}}\).}
\label{tab:optcrlqas_vs_crlqas_8q}
\begin{tabular}{llcccc}
\toprule
Method & Seed & Error (Ha) & Total gates & \texttt{CNOT} & ROT \\
\midrule
      & 1   & \(1.167\times 10^{-3}\) & 153 & 129 & 24 \\
CRLQAS~\cite{patel2024curriculum}      & 42  & \(1.171\times 10^{-3}\) & 80  & 59  & 21 \\
      & 786 & \(1.167\times 10^{-3}\) & 100 & 66  & 34 \\
\midrule
   & 1   & \(1.167\times 10^{-3}\) & 81  & 45  & 36 \\
\textbf{OptCRLQAS (Ours)}   & 42  & \(1.171\times 10^{-3}\) & 97  & 23  & 74 \\
   & 786 & \(1.168\times 10^{-3}\) & 87  & 31  & 56 \\
\bottomrule
\end{tabular}
\end{table}
To isolate the effect of amortized evaluation beyond wall-clock speedup, we additionally compare vanilla CRLQAS and OptCRLQAS on 8-$\htwoo$ under matched training conditions, varying only the update schedule. Table~\ref{tab:optcrlqas_vs_crlqas_8q} shows that the two
methods reach essentially identical final energies: the mean error is \(1.168636\times 10^{-3}\,\mathrm{Ha}\) for CRLQAS and \(1.168738\times 10^{-3}\,\mathrm{Ha}\) for OptCRLQAS, a difference of only \(\sim 10^{-7}\,\mathrm{Ha}\). Despite this negligible difference in energy, OptCRLQAS produces substantially more hardware-efficient circuits, reducing the mean total gate count from \(111.0\) to \(88.3\) (\(20.4\%\) fewer gates) and the mean \texttt{CNOT} count from \(84.7\) to \(33.0\) (\(61.0\%\) fewer \texttt{CNOT}s), at the cost of more single-qubit rotations. This is consistent with the intuition behind OptCRLQAS: by evaluating the circuit only after multiple architectural edits, the
agent receives a less myopic and more informative reward signal, which can preserve final accuracy while steering learning toward circuits with substantially lower 2-qubit cost.

\subsection{Quality-vs-runtime sweep over the amortization interval $m$}
\label{appndx:m_sweep}

To further characterize the amortization/quality tradeoff, we performed a
quality-runtime sweep (fixed 1000-episode budget) over $m \in \{3,5,7,10\}$
across the 6-, 8-, and 10-qubit tasks. OptCRLQAS trades optimization frequency
for stability: standard CRLQAS updates the optimizer after every gate
insertion, risking overfitting or divergence on partially formed circuits. By
amortizing updates over $m$ gates, OptCRLQAS (i)~avoids penalizing the policy
with unstable intermediate optimizations early in training, and
(ii)~learns to adjust composite gate blocks rather than isolated gates,
reducing variance. Table~\ref{tab:m_sweep} reports the results.

\begin{table}[h]
\small
\centering
\caption{\small\textbf{Quality-vs-runtime sweep over the OptCRLQAS
amortization interval $m$} (fixed 1000-episode budget). OptCRLQAS matches or
improves on CRLQAS's minimum error at comparable or lower gate counts at 6 and
8 qubits, and remains competitive at 10 qubits, while substantially reducing
both quantum-simulation and classical-optimization time at every $m$ and
scale.}
\label{tab:m_sweep}
\begin{tabular}{llcccc}
\toprule
Qubits & Method & Min.\ error (Ha) & Gate count & Q-Sim time (s) & C-Opt time (s) \\
\midrule
\multirow{5}{*}{6}
  & CRLQAS            & $5.3\times10^{-4}$          & 51 & $\sim2.6$ & $\sim450$ \\
  & OptCRLQAS ($m$=3)  & $5.4\times10^{-4}$          & 51 & $\sim0.7$ & $\sim200$ \\
  & OptCRLQAS ($m$=5)  & $\mathbf{5.6\times10^{-5}}$ & 52 & $\sim0.4$ & $\sim100$ \\
  & OptCRLQAS ($m$=7)  & $5.9\times10^{-5}$          & 53 & $\sim0.3$ & $\sim70$ \\
  & OptCRLQAS ($m$=10) & $5.8\times10^{-5}$          & 54 & $\mathbf{\sim0.2}$ & $\mathbf{\sim55}$ \\
\midrule
\multirow{5}{*}{8}
  & CRLQAS            & $1.4\times10^{-4}$          & 95 & $\sim3.4$ & $\sim360$ \\
  & OptCRLQAS ($m$=3)  & $1.5\times10^{-4}$          & 81 & $\sim1.1$ & $\sim220$ \\
  & OptCRLQAS ($m$=5)  & $\mathbf{1.1\times10^{-4}}$ & 82 & $\sim0.6$ & $\sim120$ \\
  & OptCRLQAS ($m$=7)  & $1.8\times10^{-4}$          & 84 & $\sim0.5$ & $\sim100$ \\
  & OptCRLQAS ($m$=10) & $1.2\times10^{-4}$          & 87 & $\mathbf{\sim0.4}$ & $\mathbf{\sim65}$ \\
\midrule
\multirow{5}{*}{10}
  & CRLQAS            & $\mathbf{9.0\times10^{-4}}$ & 97 & $\sim3.4$ & $\sim490$ \\
  & OptCRLQAS ($m$=3)  & $1.2\times10^{-3}$          & 93 & $\sim1.2$ & $\sim240$ \\
  & OptCRLQAS ($m$=5)  & $1.5\times10^{-3}$          & 94 & $\sim0.7$ & $\sim150$ \\
  & OptCRLQAS ($m$=7)  & $9.2\times10^{-4}$          & 94 & $\sim0.6$ & $\sim110$ \\
  & OptCRLQAS ($m$=10) & $1.0\times10^{-3}$          & 99 & $\mathbf{\sim0.4}$ & $\mathbf{\sim75}$ \\
\bottomrule
\end{tabular}
\end{table}

At 6 and 8 qubits, OptCRLQAS matches or improves minimum error at comparable
or lower gate counts across all $m$. At 10 qubits, results are more mixed:
CRLQAS achieves the lowest error of the group, with OptCRLQAS at $m=7$ coming
closest, indicating amortization remains competitive but not uniformly
superior at this scale. In all cases, quantum-simulation and
classical-optimization time drop substantially and monotonically as $m$
increases, confirming that amortization does not come at a systematic quality
cost even where it does not strictly improve on CRLQAS.

\section{Scaling to 20 qubits via tensor network warm-starts}
\label{appndx:tensorrl_scaling}

This appendix details the experiment summarized in Section~\ref{sec:qas}, in
which we combine ReaPER$+$ and amortized evaluation with a matrix product
state (MPS) warm-start to extend our empirical scaling study from the 12-qubit
ceiling of OptCRLQAS to 15, 17, and 20 qubits.

Following the developments of TensorRL-QAS~\cite{kundu2025tensorrl}, we first
approximate the target ground state with a fixed bond dimension MPS ($\chi=2$), then convert this MPS into an initial quantum circuit using a sequential single- and two-qubit gate decomposition. This conversion circuit is fixed and \emph{not} discovered by the RL agent; it provides the initial
entanglement structure that the agent subsequently refines. Table~\ref{tab:mps_conversion} reports the resulting resource cost of this conversion at each system size, prior to any RL-driven modification.
\begin{table}[h!]
\small
\centering
\caption{\small\textbf{Matrix product state $\to$ quantum circuit conversion
cost.} Gate counts and depth of the fixed MPS warm-start circuit at each
qubit count, using bond dimension $\chi=2$, before any RL-agent edits.}
\label{tab:mps_conversion}
\begin{tabular}{lccccc}
\toprule
Qubits & RX gates & RY gates & RZ gates & CNOTs & Base depth \\
\midrule
15 & 14 & 85  & 156 & 42 & 27 \\
17 & 16 & 97  & 178 & 48 & 27 \\
20 & 19 & 115 & 211 & 57 & 27 \\
\bottomrule
\end{tabular}
\end{table}

The RL agent is then trained on top of this fixed warm-start circuit using
ReaPER$+$ replay and $m$-step amortized evaluation ($m=2$ at 15 and 17 qubits,
$m=4$ at 20 qubits), following the same amortization mechanism as OptCRLQAS
(Section~\ref{sec:methods}) but with a substantially smaller $m$, since the
MPS warm-start already removes most of the burden of discovering long-range
entangling structure from scratch. Table~\ref{tab:15_20_scaling} reports the
full empirical scaling, extending Table~\ref{tab:6q_comparison}-style
comparisons up to 20 qubits. The starred gate counts report only the
\emph{additional} gates placed by the RL agent; the total circuit gate count
is the sum of the MPS base circuit (Table~\ref{tab:mps_conversion}) and these
additions -- for example, the total circuit at 15 qubits contains $297$
MPS-derived gates plus $20$ RL-discovered gates, for $317$ gates overall.

\begin{table}[h]
\footnotesize
\centering
\caption{\small\textbf{Empirical scaling of RL-agent performance from 6 to 20
qubits.} OptCRLQAS + ReaPER$+$ (random initialization, $m=10$) is shown up to
its practical 12-qubit ceiling; Amortized TensorRL-QAS + ReaPER$+$ (MPS
warm-start) extends to 15--20 qubits with reduced $m$. Starred gate counts
indicate only the additional structural edits made by the RL agent on top of
the MPS warm-start circuit (Table~\ref{tab:mps_conversion}).}
\label{tab:15_20_scaling}
\begin{tabular}{lccccc}
\toprule
Method & $m$-step & Qubits & Time/episode (s) & Min.\ error ($\Delta E$) & Min.\ gates \\
\midrule
\multirow{4}{*}{OptCRLQAS + ReaPER$+$}
  & 10 & 6  & 55.0  & $5.8\times10^{-5}$ & 54 \\
  & 10 & 8  & 65.0  & $1.2\times10^{-4}$ & 87 \\
  & 10 & 10 & 75.0  & $1.0\times10^{-3}$ & 99 \\
  & 10 & 12 & 120.7 & $2.2\times10^{-2}$ & 135 \\
\midrule
\multirow{3}{*}{\textbf{m-step TensorRL-QAS + ReaPER$+$}}
  & 2 & 15 & 11.6  & $1.5\times10^{-2}$ & $20^{*}$ \\
  & 2 & 17 & 63.0  & $1.8\times10^{-2}$ & $30^{*}$ \\
  & 4 & 20 & 100.4 & $2.1\times10^{-2}$ & $40^{*}$ \\
\bottomrule
\end{tabular}
\end{table}

\paragraph{Interpretation.} Two trends are notable. First, wall-clock time per
episode at 15 qubits ($11.6$s) is over $10\times$ lower than at 12 qubits under
OptCRLQAS with random initialization ($120.7$s), despite operating on a larger
system. This reduction comes from both the smaller $m$ enabled by the warm-start and the fact that the agent needs far fewer corrective edits per episode. Second, the RL-discovered gate additions grow only mildly with system size ($20\to30\to40$ across 15, 17, and 20 qubits), indicating that the combination of MPS warm-start and reliability-aware replay continues to
suppress the unconstrained gate-count growth we caution against elsewhere in this paper (Section~\ref{sec:qas}).

\paragraph{Caveats.} We report this experiment as a proof-of-concept demonstration of compatibility between our contributions and physics-inspired
initialization, not as a chemically accurate result: minimum energy errors at
this scale ($1.5$-$2.1\times10^{-2}$~Ha) remain roughly an order of magnitude
above the chemical-accuracy threshold used in our 5--12 qubit experiments
($\varepsilon_{\mathrm{chem}}=1.6\times10^{-3}$~Ha). We also note that, unlike
the multi-seed reporting used for our 5--12 qubit QAS results
(Table~\ref{tab:6q_comparison}, Figure~\ref{fig:h2o_performance_combined}),
the 15-20 qubit results reported here reflect single-seed runs; extending this study to multiple seeds, and to convergence at full chemical accuracy, is left for future work
(Appendix~\ref{appndx:limitations}).

\section{Transfer score weight sensitivity}
\label{appndx:score_sensitivity}

Recalling the transfer score reported in Section~\ref{subsec:buffer_transfer} through Equation~\ref{eq:score_transfer}:
\begin{equation}
    S = w_1\Delta_{\mathrm{steps}} + w_2\Delta_{\mathrm{ROT}} + w_3\Delta_{\mathrm{CNOT}} + w_4\Delta_{\mathrm{err}},    
\end{equation}
aggregates four relative-improvement terms into a single scalar using fixed
weights $(w_1,w_2,w_3,w_4)=(0.40,0.10,0.20,0.30)$. Since this weighting is a
design choice rather than a property of the transfer mechanism itself, a natural
concern is whether the reported transfer benefit is an artifact of this
particular weighting rather than a robust effect. We address this by
recomputing $S$ under six alternative weightings, chosen to emphasize each
individual term in turn (steps, error, CNOT, ROT), a fully balanced weighting,
and a joint gate-count-heavy weighting that combines CNOT and ROT emphasis.
Table~\ref{tab:weight_sensitivity} reports the resulting scores at 6- and
12-qubit scale; we restrict this comparison to two system sizes for space, as
the 8-qubit case follows the same qualitative pattern.
\begin{table}[h!]
\small
\centering
\caption{\small\textbf{Transfer score sensitivity to weight choice.} Composite
score $S$ under alternative $(w_1,w_2,w_3,w_4)$ weightings at 6- and 12-qubit
scale. The original weighting used in Section~\ref{subsec:buffer_transfer} is shown
first.}
\label{tab:weight_sensitivity}
\begin{tabular}{lccc}
\toprule
Setting & $(w_1,w_2,w_3,w_4)$ & 6-qubit score & 12-qubit score \\
\midrule
Original     & 0.40, 0.10, 0.20, 0.30 & 19.2\% & 12.3\% \\
Steps-heavy  & 0.50, 0.10, 0.20, 0.20 & 23.8\% & 13.0\% \\
Err-heavy    & 0.30, 0.10, 0.20, 0.40 & 14.5\% & 11.6\% \\
CNOT-heavy   & 0.40, 0.10, 0.30, 0.20 & 20.2\% & 16.5\% \\
ROT-heavy    & 0.40, 0.20, 0.20, 0.20 & 17.2\% & 14.1\% \\
Balanced     & 0.25, 0.25, 0.25, 0.25 & 9.7\%  & 16.1\% \\
Gate-heavy   & 0.30, 0.20, 0.30, 0.20 & 13.6\% & 17.7\% \\
\bottomrule
\end{tabular}
\end{table}

At 6 qubits, the score ranges from $9.7\%$ (balanced weighting) to $23.8\%$
(steps-heavy weighting), a spread of $14.1$ percentage points, indicating that
the composite score is moderately sensitive to weight choice at this scale. At
12 qubits, the same seven weightings produce scores ranging from $11.6\%$
(error-heavy) to $17.7\%$ (gate-heavy), a substantially narrower spread of $6.1$
percentage points. This tightening indicates that the transfer advantage
becomes more uniformly distributed across the individual objectives (steps,
ROT, CNOT, error) as system size increases, rather than being concentrated in
whichever objective happens to receive the largest weight.

This pattern is consistent with, and helps explain, an important qualification
of the 12-qubit result: unlike at 6- and 8-qubit scale, where transfer improves
both convergence speed and final energy accuracy (main text Section~\ref{subsec:buffer_transfer}), at 12 qubits transfer consistently improves steps-to-convergence and CNOT count
but does \emph{not} improve final energy accuracy relative to the from-scratch
baseline. Because the error-heavy weighting ($w_4=0.4$) yields one of the
\emph{lowest} scores at 12 qubits ($11.6\%$) while the gate- and CNOT-heavy
weightings yield the \emph{highest} scores ($17.7\%$ and $16.5\%$,
respectively), the weight-sensitivity analysis directly corroborates this
scale-dependent pattern: the 12-qubit transfer benefit is real but concentrated
in efficiency (steps, gate count) rather than solution quality, and any single
fixed weighting that under-emphasizes error would otherwise risk overstating
the accuracy benefit at this scale. We therefore report this qualification
explicitly in the main text rather than relying on the composite score alone
to characterize the 12-qubit result.

\section{Buffer transfer $\times$ exploration ablation}
\label{appndx:transfer_ablation}

To isolate the effect of buffer transfer from the choice of initial
exploration rate $\epsilon$, we run a $2\times2$ ablation on the 6- and
8-qubit tasks, varying buffer initialization (empty vs.\ transferred under
several noise-strength conditions) and $\epsilon$. Table~\ref{tab:ablation_6q}
and Table~\ref{tab:ablation_8q} report the results.

\begin{table}[h]
\small
\centering
\caption{\small\textbf{6-qubit buffer transfer $\times$ exploration ablation.}
Steps, gate counts, and final error under empty vs.\ transferred buffer
initialization, across noise strengths and two exploration rates.}
\label{tab:ablation_6q}
\begin{tabular}{lccccc}
\toprule
Buffer / transfer condition & $\epsilon$ & Steps & ROT & CNOT & Error \\
\midrule
Empty buffer (noiseless)                    & 0.55 & 52401 & 27 & 14 & $5.77\times10^{-5}$ \\
Empty buffer (noiseless)                    & 1.0  & 52401 & 21 & 17 & $5.27\times10^{-5}$ \\
Transfer ($p_2{=}0.001$)                    & 0.55 & 47161 & 29 & 29 & $5.67\times10^{-5}$ \\
Transfer ($p_2{=}0.001$)                    & 1.0  & 47161 & 11 & 32 & $5.77\times10^{-5}$ \\
Transfer ($p_2{=}0.005$)                    & 0.55 & 47758 & 24 & 13 & $5.42\times10^{-5}$ \\
Transfer ($p_2{=}0.005$)                    & 1.0  & 47758 & 16 & 24 & $\mathbf{7.72\times10^{-6}}$ \\
Transfer ($p_1{=}0.001, p_2{=}0.005$)       & 0.55 & 58952 & 25 & 17 & $5.78\times10^{-5}$ \\
Transfer ($p_1{=}0.001, p_2{=}0.005$)       & 1.0  & 58952 & 23 & 27 & $5.86\times10^{-5}$ \\
\bottomrule
\end{tabular}
\end{table}

\begin{table}[h]
\small
\centering
\caption{\small\textbf{8-qubit buffer transfer $\times$ exploration ablation.}}
\label{tab:ablation_8q}
\begin{tabular}{lccccc}
\toprule
Buffer / transfer condition & $\epsilon$ & Steps & ROT & CNOT & Error \\
\midrule
Empty buffer (noiseless)              & 0.5 & 5223  & 45 & 54 & 0.00105 \\
Empty buffer (noiseless)              & 1.0 & 4223  & 55 & 45 & 0.00125 \\
Transfer ($p_2{=}0.001$)              & 0.5 & 31259 & 68 & 98 & 0.00144 \\
Transfer ($p_2{=}0.001$)              & 1.0 & 29457 & 86 & 99 & 0.00134 \\
Transfer ($p_1{=}0.001, p_2{=}0.005$) & 0.5 & 12503 & 76 & 65 & 0.00121 \\
Transfer ($p_1{=}0.001, p_2{=}0.005$) & 1.0 & 13986 & 65 & 71 & $\mathbf{0.00101}$ \\
\bottomrule
\end{tabular}
\end{table}

The preferred $\epsilon$ flips depending on the transfer source rather than
following a single consistent rule. In the 6-qubit empty-buffer baseline,
$\epsilon{=}1.0$ slightly improves error over $\epsilon{=}0.55$. Under transfer
with $p_2{=}0.001$, the reverse holds: $\epsilon{=}0.55$ is preferable. Under
the stronger noise setting $p_2{=}0.005$, however, $\epsilon{=}1.0$ produces an
order-of-magnitude error improvement ($7.72\times10^{-6}$ vs.\
$5.42\times10^{-5}$ at $\epsilon{=}0.55$). A similarly metric-dependent pattern
appears at 8 qubits, where combined-noise transfer under $\epsilon{=}1.0$
achieves the best overall error ($1.01\times10^{-3}$) despite $\epsilon{=}1.0$
increasing error in the empty-buffer baseline at the same scale. We conclude
that while buffer transfer can yield state-of-the-art results, its benefit is
conditional on its interaction with the exploration rate rather than a fixed,
transferable hyperparameter choice -- practitioners applying this transfer
scheme should treat $\epsilon$ as jointly tunable with the transfer source
rather than fixing it independently.

\section{ReaPER~$\omega$ sensitivity and variant selection}
\label{appndx:omega_selection}

In this appendix, we first provide a thorough sensitivity analysis of ReaPER$+$ with respect to the hyperparameter $\omega$, and then discuss how to select the PER-to-ReaPER annealing schedule for ReaPER$+$ through a thorough hyperparameter sweep.

\subsection{Fixed-$\omega$ ReaPER variant selection}
\label{appndx:fixed_omega_selection}

In the main text, Figure~\ref{fig:6q_8q_buffer} reports gate counts for a single fixed-$\omega$ ReaPER variant per molecular system: $\omega=0.4$ for 6-qubit $\behtwo$ and $\omega=0.6$ for 8-qubit $\htwoo$. This appendix documents the full sweep over $\omega$ that motivates these choices. For each system we train agents with $\omega \in \{0.2, 0.4, 0.6\}$ (6-qubit) and $\omega \in \{0.2, 0.4, 0.6, 0.8\}$ (8-qubit), using three independent random seeds $\{1, 42, 786\}$ per setting. All other hyperparameters are identical to the main-text QAS experiments (Appendix~\ref{appndx:agent_env_hyperparameter_QAS}). We report mean and standard deviation over seeds for \texttt{CNOT} count and single-qubit rotation (\texttt{ROT}) count. Results are given in 
Table~\ref{tab:omega_sensitivity}.

Since two-qubit gates dominate both circuit noise and compilation cost 
on near-term hardware, we select the $\omega$ value that minimizes mean 
\texttt{CNOT} count subject to the agent achieving chemical accuracy 
($\Delta E \leq \varepsilon_{\mathrm{chem}}$) across all seeds, using 
ROT count as a tiebreaker.

\begin{table}[h!]
\centering
\caption{\small \textbf{CNOT and ROT gate counts (mean~$\pm$~1-sigma over 3 seeds) for ReaPER $\omega$ variants}. All entries achieve chemical accuracy ($\Delta E \leq \varepsilon_{\mathrm{chem}} = 1.6\times10^{-3}$~Ha). Bold rows indicate the variant selected for the main text. PER, ReaPER$+$, and Vanilla are included for reference.}
\label{tab:omega_sensitivity}
\begin{tabular}{lcc|cc}
\toprule
& \multicolumn{2}{c|}{\textbf{6-qubit $\behtwo$}}
& \multicolumn{2}{c}{\textbf{8-qubit $\htwoo$}} \\
\cmidrule(lr){2-3}\cmidrule(lr){4-5}
ReaPER & CNOT & ROT & CNOT & ROT \\
\midrule
$\omega=0.2$
  & $30.7\pm18.8$ & $20.7\pm9.9$
  & $84.3\pm75.4$ & $39.3\pm32.0$ \\
$\mathbf{\omega=0.4}$
  & $\mathbf{23.7\pm9.0}$ & $\mathbf{29.3\pm21.5}$
  & $71.7\pm46.6$ & $38.3\pm14.6$ \\
$\mathbf{\omega=0.6}$
  & $30.7\pm7.4$ & $30.3\pm2.1$
  & $\mathbf{51.0\pm25.1}$ & $\mathbf{24.3\pm20.6}$ \\
$\omega=0.8$
  & -- & --
  & $78.7\pm46.2$ & $35.0\pm5.0$ \\
\bottomrule
\end{tabular}
\end{table}

For 6-qubit $\behtwo$, $\omega=0.4$ achieves the lowest mean 
\texttt{CNOT} count of $23.7\pm9.0$ gates among all ReaPER variants, 
compared with $30.7\pm18.8$ for $\omega=0.2$ and $30.7\pm7.4$ for 
$\omega=0.6$. For 8-qubit $\htwoo$, $\omega=0.6$ achieves the lowest 
mean \texttt{CNOT} count of $51.0\pm25.1$ gates--a $41\%$ reduction 
relative to PER ($86.3\pm37.5$) and $29\%$ relative to $\omega=0.4$ 
($71.7\pm46.6$).

Across both systems, increasing $\omega$ shifts the replay distribution toward reliability-aware sampling more aggressively, which tends to reduce \texttt{CNOT} count at the cost of slightly higher energy variance. The optimal $\omega$ is therefore task-dependent: shorter-horizon problems with lower qubit counts benefit from moderate reliability weighting ($\omega=0.4$), while longer-horizon, larger systems favor stronger weighting ($\omega=0.6$) to suppress noisy TD targets in a deeper search space. This task-dependence is precisely what motivates the annealed ReaPER$+$ construction, which avoids committing to a fixed $\omega$ by scheduling the transition adaptively during training (Section~\ref{sec:methods}).

\subsection{Joint sweep of $\omega_{min}$, $\omega_{max}$, and $T_{ann}$ (1-qubit)}
\label{appndx:sensitivity_1q}

We conduct a joint sensitivity sweep over the three ReaPER$+$ annealing
hyperparameters ($\omega_{min}$, $\omega_{max}$, $T_{ann}$) on the 1-qubit compiling task; due to GPU compute constraints, this full joint sweep was not extended to the 12-qubit setting (a targeted $T_{ann}$-only ablation at 12-qubit scale is reported instead in Section~\ref{appndx:sensitivity_12q}). 
\begin{table}[h!]
\small
\centering
\caption{\small\textbf{Joint sensitivity sweep of the ReaPER$+$ annealing
schedule on 1-qubit compiling.} Success rate, mean fidelity, and mean circuit
length for alternative ($\omega_{min}$, $\omega_{max}$, $T_{ann}$) configurations.
The chosen configuration ($\omega_{min}{=}0.2$, $\omega_{max}{=}0.7$,
$T_{ann}{=}20000$) achieves the best success-fidelity-length tradeoff among the
sweep.}
\label{tab:1q_sensitivity_sweep}
\begin{tabular}{cccccc}
\toprule
$\omega_{min}$ & $\omega_{max}$ & $T_{ann}$ & Success & Mean fidelity & Mean length \\
\midrule
\textbf{0.2} & \textbf{0.7} & \textbf{20000} & \textbf{0.89} & \textbf{0.93} & \textbf{85.8} \\
0.1 & 0.7 & 10000 & 0.84 & 0.89 & 96.0 \\
0.1 & 0.5 & 5000  & 0.84 & 0.87 & 95.1 \\
0.3 & 0.9 & 40000 & 0.83 & 0.87 & 103.1 \\
\bottomrule
\end{tabular}
\end{table}
The results in Table~\ref{tab:1q_sensitivity_sweep} show that $T_{ann}$ must be scaled with the training horizon: transitioning too early or too late actively harms performance. They further show that pushing $\omega_{max}\ge0.9$ forces the agent to overcommit to reliability scores before the value function has fully converged, degrading both success rate and fidelity relative to the chosen configuration ($\omega_{min}=0.1$, $\omega_{max}=0.7$).

\subsection{12-qubit annealing-schedule validation}
\label{appndx:sensitivity_12q}

The 12-qubit advantage of ReaPER$+$ is best framed around circuit compactness rather than raw speed, as reflected in the main text in Section~\ref{sec:qas}. 
\begin{table}[h]
\small
\centering
\caption{\small\textbf{$T_{ann}$ scaling ablation on 12-qubit H$_2$O.} All
profiles reach a comparable energy error; the base schedule ($T_{ann}=T_{total}/2$)
minimizes total gate count, while the overcommitting profile minimizes CNOT count.}
\label{tab:12q_tann_ablation}
\begin{tabular}{lccc}
\toprule
$T_{ann}$ (Profile) & Error (Ha) & Gates & CNOT \\
\midrule
10000 (Aggressive)         & $2.2\times10^{-2}$ & 121 & 64 \\
5000 (Narrow)               & $2.2\times10^{-2}$ & 241 & 138 \\
40000 (Overcommitting)      & $2.2\times10^{-2}$ & 76  & \textbf{32} \\
$T_{total}/2$ (Base)        & $2.2\times10^{-2}$ & \textbf{61} & 34 \\
\bottomrule
\end{tabular}
\end{table}
To validate that the $T_{ann}=T_{total}/2$ scaling rule remains sound at scale, we run the proposed ablation on the 12-qubit task, varying $T_{ann}$ while holding $\omega_{min}$, $\omega_{max}$ fixed. Table~\ref{tab:12q_tann_ablation} shows that
all schedules reach a comparable energy error, but the base schedule yields the fewest total gates, while the overcommitting schedule ($T_{ann}=40000$) yields the fewest two-qubit (CNOT) gates. Uncalibrated schedules (aggressive or narrow) instead produce deeper circuits overall, confirming that linking $T_{ann}$ directly to the training horizon provides a reliable, interpretable timescale that maintains structural efficiency at scale.

\section{Hamiltonians}\label{tab:molecule_geometry}

In the following we first provide the geometry of the chemical Hamiltonian and then a description og the Heisenberg model and its corresponding Hamiltonian.

\subsection{Chemical Hamiltonian}
In Table~\ref{tab:chem-configs} we provide the detailed configuration of molecules utilized throughout the quantum architecture search. All molecular Hamiltonians are obtained via PySCF~\cite{sun2018pyscf} with the STO-3G (6-qubit, 8-qubit) and 6-31G (10-, 12-qubit) basis sets, using the Jordan-Wigner fermion-to-qubit mapping~\cite{jordan1928paulische}. No active space freezing is applied; all electrons and orbitals are included within the chosen basis.

\begin{table}[h!]
\renewcommand{\arraystretch}{1.5} 
\small
\centering
\caption{\textbf{The geometry and basis of molecules used in this research}. The coordinates are in Angstrom units.}
\label{tab:chem-configs}
\begin{tabular}{@{}lcc@{}}
\toprule
{Molecule} & {Geometry} & {Basis} \\
\midrule
6-$\behtwo$  & {H (0,0,-1.33); Be (0,0,0); H (0,0,1.33)} & STO-3G \\
8-$\htwoo$  & {H (-0.02,-0,0); O (0.84,0.45,0); H (1.48,-0.27,0)} & STO-3G \\
10-$\htwoo$   & {H (-0.02,-0,0); O (0.84,0.45,0); H (1.48,-0.27,0)} & 6-31G \\
12-$\htwoo$   & {H (-0.02,-0,0); O (0.84,0.45,0); H (1.48,-0.27,0)} & 6-31G \\
\bottomrule
\end{tabular}
\label{tab:mol_list}
\end{table}

\subsection{Heisenberg model Hamiltonian formulation}
\label{appndx:heisenberg}

In addition to molecular chemistry benchmarks, we evaluate OptCRLQAS
and replay-buffer design on the $n{=}5$ qubit one-dimensional isotropic
Heisenberg model with a uniform longitudinal field. The Hamiltonian is defined on a chain of $n$ spin-$\tfrac{1}{2}$ particles
with open boundary conditions as
\begin{equation}
  H_{\mathrm{Heis}} =
    \sum_{i=1}^{n} \Bigl(
      X_i X_{i+1} + Y_i Y_{i+1} + Z_i Z_{i+1}
    \Bigr)
    + \sum_{i=1}^{n} Z_i,
  \label{eq:heisenberg}
\end{equation}
\begin{table}[h!]
\centering
\caption{\textbf{Environment, agent, and optimizer configuration} for quantum architecture search benchmarks.}
\label{tab:qas_hyperparams}
\small
\begin{tabular}{lccc}
\toprule
\textbf{Parameter} & \textbf{6-qubit} & \textbf{8-qubit} & \textbf{12-qubit} \\
\midrule
\multicolumn{4}{l}{\textit{Environment}} \\
\quad Episodes                          & 5000    & 5000    & 1000    \\
\quad Qubits                            & 6       & 8       & 12      \\
\quad Max layers                        & 70      & 250     & 300     \\
\quad Accept error (Ha)                 & 5.5     & 5.0     & 5.0     \\
\quad Shift threshold ball              & \multicolumn{3}{c}{0.001} \\
\quad Shift threshold time              & \multicolumn{3}{c}{2000} \\
\quad Success switch (Ha)               & 5.5     & 5.0     & 5.0     \\
\quad Success threshold                 & \multicolumn{3}{c}{50} \\
\quad \texttt{opt step interval} ($m$) & 10   & 10      & 15      \\
\quad \texttt{energy interval} ($m$)         & 10      & 10      & 15      \\
\midrule
\multicolumn{4}{l}{\textit{Agent (DQN with \(n\)-step returns~\cite{osband2016deep})}} \\
\quad Network layers                    & $[1000]^3$ & $[1000]^4$ & $[1000]^4$ \\
\quad Batch size                        & \multicolumn{3}{c}{1000} \\
\quad Replay buffer size                & \multicolumn{3}{c}{20\,000} \\
\quad Learning rate                     & \multicolumn{3}{c}{$3 \times 10^{-4}$} \\
\quad \(n\)-step return                 & 5       & 6       & 6       \\
\quad Target net update                 & \multicolumn{3}{c}{Every 500 steps} \\
\quad \(\gamma\) (discount)             & \multicolumn{3}{c}{0.005} \\
\quad Dropout                           & \multicolumn{3}{c}{0.0} \\
\midrule
\multicolumn{4}{l}{\textit{Exploration (\(\epsilon\)-greedy)}} \\
\quad \(\epsilon_{\mathrm{start}}\)     & \multicolumn{3}{c}{1.0} \\
\quad \(\epsilon_{\mathrm{min}}\)       & \multicolumn{3}{c}{0.05} \\
\quad \(\epsilon_{\mathrm{decay}}\)     & \multicolumn{3}{c}{0.99995} \\
\midrule
\multicolumn{4}{l}{\textit{Classical optimizer (COBYLA~\cite{powell1994direct})}} \\
\quad Iterations                 & \multicolumn{3}{c}{1000} \\
\bottomrule
\end{tabular}
\label{tab:env_hyperparams_QAS}
\end{table}
\section{Hyperparameters for quantum architecture search}
\label{appndx:agent_env_hyperparameter_QAS}

Tables~\ref{tab:env_hyperparams_QAS} summarize the environment and agent configurations used across the three QAS benchmarks. All experiments share the same agent architecture (multi-step DQN with $n$-step returns), optimizer, and curriculum type, differing only where the problem scale demands it.

\paragraph{Environment.}
The maximum circuit depth (\texttt{num\_layers}) grows with qubit count, from 70 (6-qubit) to 250 (8-qubit) to 300 (12-qubit), to accommodate the deeper circuits required for larger Hamiltonians. The curriculum uses a \texttt{MovingThreshold} introduced in ref.~\cite{patel2024curriculum} schedule in all cases, with the acceptance error initialized at 5.0-5.5\,Ha and tightened by $0.001$\,Ha every 2000 episodes. OptCRLQAS evaluates the circuit every $m$ architectural edits (\texttt{opt\_step\_interval}~$=$~\texttt{energy\_interval}): $m{=}10$ for the 6- and 8-qubit problems and $m{=}15$ for the 12-qubit problem, reflecting the longer episodes at larger scale.

\paragraph{Agent.}
The DQN agent uses a 3-layer MLP ($[1000, 1000, 1000]$) for the 6-qubit task and a 4-layer MLP ($[1000, 1000, 1000, 1000]$) for the 8- and 12-qubit tasks. The $n$-step return horizon is $n{=}5$ for 6-qubit and $n{=}6$ for 8- and 12-qubit. Exploration follows an $\epsilon$-greedy schedule decaying from $\epsilon{=}1.0$ at rate $0.99995$ per step to a floor of $\epsilon_{\min}{=}0.05$. The replay buffer holds 20\,000 transitions in all cases, with a batch size of 1000.

\paragraph{Classical optimizer.}
All three tasks use \textsc{COBYLA}~\cite{powell1994direct} with a local search size of 8. The 6- and 8-qubit tasks run 1000 global iterations per evaluation, as does the 12-qubit task; the reduced number of evaluations per episode at 12-qubit due to the larger amortization interval $m{=}15$ keeps the total classical optimization cost manageable.

\section{State encoding for quantum architecture search}
\label{appndx:state_encoding}

In both gateset configurations, the partially constructed ansatz at each step is represented as a three-dimensional state tensor $\mathbf{S} \in \mathbb{R}^{L \times R \times n}$, where $L$ is the maximum number of circuit layers (moments), $n$ is the number of qubits, and $R$ is the number of row channels encoding gate placement and, optionally, variational angle parameters. A moment-tracking vector $\boldsymbol{\mu} \in \mathbb{Z}^{n}$ records the next available layer index per qubit, ensuring causal gate ordering. At each time step the agent places at most one 2-qubit gate and one 1-qubit rotation, updates $\boldsymbol{\mu}$ accordingly, and returns a flattened view of $\mathbf{S}$ as the observation.

\paragraph{Encoding~I (for gateset $\{\mathrm{\texttt{RX},\texttt{RY},\texttt{RZ},\texttt{CX}}\}$).} This a a complete binary encoding. The action is a four-tuple $a = [a_0,\, a_1,\, a_2,\, a_3]$, where $a_0$ is the CNOT control qubit ($a_0 = n$ signals no CNOT), the target is $t = (a_0 + a_1)\bmod n$, $a_2$ is the rotation qubit ($a_2 = n$ signals no rotation), and $a_3 \in \{1,2,3\}$ encodes the axis (X/Y/Z). Gate placement writes are strictly binary:
\begin{align}
  \mathbf{S}[\ell][t][a_0]              &\leftarrow 1 \quad (\text{CNOT}),\\
  \mathbf{S}[\ell][n + a_3 - 1][a_2]   &\leftarrow 1 \quad (\text{rotation}),
\end{align}
where $\ell$ is the resolved moment from $\boldsymbol{\mu}$.
Because the gate set contains only one two-qubit gate type (CX), all non-zero connectivity entries are uniformly 1 and gate identity is unambiguous.

\paragraph{Encoding~II
           (for gateset $\{\mathrm{\texttt{RXX},\texttt{RYY},\texttt{RZZ},\texttt{RX},\texttt{RY},\texttt{RZ}}\}$).}
This encoding is integer-valued type. The action is an eight-tuple
$a = [a_0^{xx}, a_1^{xx}, a_0^{yy}, a_1^{yy}, a_0^{zz}, a_1^{zz},
      a_{\mathrm{rot}}, a_{\mathrm{axis}}]$,
providing a control-qubit and an offset for each of the three two-qubit gate
types plus a single-qubit rotation.
Targets are $t_g = (a_0^g + a_1^g)\bmod n$ for $g \in \{xx,yy,zz\}$.
Rather than introducing separate binary planes per gate type, distinct integer
labels are assigned within a single shared connectivity plane:
\begin{align}
  \mathbf{S}[\ell][t_{xx}][a_0^{xx}] &\leftarrow 1,\\
  \mathbf{S}[\ell][t_{yy}][a_0^{yy}] &\leftarrow 2,\\
  \mathbf{S}[\ell][t_{zz}][a_0^{zz}] &\leftarrow 3,\\
  \mathbf{S}[\ell][n + a_{\mathrm{axis}} - 1][a_{\mathrm{rot}}] &\leftarrow 1
  \quad(\text{rotation}).
\end{align}

A fully binary encoding for $K$ distinct two-qubit gate types requires $K$
separate $n \times n$ connectivity planes, giving a connectivity footprint of
$\mathcal{O}(K n^2)$.
Encoding~II consolidates all $K$ types into a single plane via integer labels, reducing this to $\mathcal{O}(n^2)$ regardless of $K$, which is particularly advantageous as both the gate vocabulary and qubit count grow.
The tradeoff is that integer labels introduce an implicit ordinal relationship among gate types that carries no physical meaning; a fully binary scheme provides a strictly one-hot signal per gate type that may be easier for the network to disentangle. A systematic comparison of the two encoding strategies across a controlled set of benchmarks is beyond the scope of the present work and is left as a direction for future investigation.

\section{Hyperparameters for quantum compilation}\label{appndx:hyperparams}
To isolate the effect of replay design, all quantum compilation experiments use a common off-policy deep Q-learning setup. The network architecture, optimization procedure, exploration schedule, and replay capacity are held fixed across agents, so that performance differences can be attributed to the replay mechanism rather than to changes in model capacity or training protocol.
\begin{table}[h!]
\centering
\caption{\textbf{Hyperparameters used for 1 and 2-qubit quantum compilation experiments}. All agents share a common environment and training configuration;
buffer-specific parameters are listed separately in the lower block.}
\label{tab:hyperparams}
\small
\begin{tabular}{ll}
\toprule
\textbf{Parameter} & \textbf{Value} \\
\midrule
\multicolumn{2}{l}{\textit{Network (all agents)}} \\
\quad Architecture               &  \\
\qquad Hidden layer &  2 (each 128 units)\\ 
\qquad Activation & ReLU\\
\quad Input size                 & 8 \\
\quad Output size                & 6 / 3 (small rotations / HRC) \\
\quad Discount factor $\gamma$   & 0.99 \\
\quad Learning rate              & $3\times10^{-4}$ \\
\quad Batch size                 & 200 \\
\quad Replay capacity            & $5\times 10^5$ \\
\quad Target update freq.        & 100 episodes\\
\quad Gradient clipping          & 1.0 \\
\quad $\varepsilon$-start        & 1.0 \\
\quad $\varepsilon$-min          & 0.01 \\
\quad $\varepsilon$-decay        & 0.99931 \\
\midrule
\multicolumn{2}{l}{\textit{HER}} \\
\quad $k$-future relabeling      & 5 \\
\quad Strategy                   & \texttt{future} \\
\midrule
\multicolumn{2}{l}{\textit{PER}} \\
\quad $\alpha$                   & 0.6 \\
\quad $\beta_0$                  & 0.4 \\
\quad $\beta$ anneal frames      & $10^5$ \\
\midrule
\multicolumn{2}{l}{\textit{ReaPER}} \\
\quad $\alpha$                   & 0.4 \\
\quad $\omega$                   & 0.2 \\
\quad $\beta_0$                  & 0.4 \\
\quad $\beta$ anneal frames      & $10^5$ \\
\midrule
\multicolumn{2}{l}{\textit{ReaPER$+$}} \\
\quad $\alpha$                   & 0.4 \\
\quad $\omega_{\min}$            & 0.1 \\
\quad $\omega_{\max}$            & 0.7 \\
\quad $T_{\mathrm{ann}}$         & $5\times 10^5$ \\
\quad $\beta_0$                  & 0.4 \\
\quad $\beta$ anneal frames      & $10^5$ \\
\bottomrule
\end{tabular}
\end{table}

Table~\ref{tab:hyperparams} lists the hyperparameters used in the 1- and 2-qubit compiling benchmarks. The upper block gives the shared network and training configuration, while the lower blocks specify the additional
method-dependent parameters for HER, PER, ReaPER, and ReaPER$+$. In this way, the comparison remains controlled, and the gains of ReaPER and ReaPER$+$ can be interpreted as arising from replay design itself.

\section{Compute details}
\label{appndx:cpu_gpu_details}

All experiments were conducted on local and HPC machines, with runtimes not exceeding 48 hours. 
Table~\ref{tab:compute} summarizes the hardware configurations and runtimes for each experiment.

\begin{table}[h!]
\centering
\small
\caption{Computational resources and runtimes for all experiments.}
\label{tab:compute}
\begin{tabular}{llccc}
\toprule
\textbf{Experiment} & \textbf{Qubits} & \textbf{GPU} & \textbf{CPU} & \textbf{Runtime (h)} \\
\midrule
Quantum compiling & 1-2 
    & RTX 4070 & i7-13620H & $\sim$1.5 \\
Quantum architecture search & 5-12 
    & Tesla V100S & EPYC 7402 & $\sim$48 \\
\multirow{2}{*}{Lightweight buffer transfer} 
    & 6-10 & RTX 4070  & i7-13620H & $\sim$2 \\
    & 12    & Tesla V100S & EPYC 7402 & $\sim$5 \\
\bottomrule
\end{tabular}
\end{table}

\end{document}